\documentclass[onecolumn,floatfix,superscriptaddress,a4paper,
               showpacs,showkeys,nofootinbib,preprint]{revtex4}

\usepackage{epsfig}
\usepackage{latexsym}
\usepackage{xspace}
\usepackage{hyperref}
\usepackage[latin2]{inputenc}
\usepackage{indentfirst}
\usepackage{enumerate}

\usepackage{amsmath}
\usepackage{amssymb}
\usepackage[english]{babel}
\usepackage{url}

\newcommand{\bs}{\boldsymbol}

\newcommand*{\dis}{\displaystyle}

\begin{document}

\title{\bf Single-Particle Spectra in Relativistic Heavy-Ion Collisions
\newline
Within the Thermal Quantum Field Theory}
\author{D. Anchishkin}
\affiliation{ Bogolyubov Institute for Theoretical Physics, 03143 Kyiv, Ukraine}
\affiliation{ Taras Shevchenko National University of Kyiv, 03022 Kyiv, Ukraine}
\affiliation{Frankfurt Institute for Advanced Studies,
60438 Frankfurt am Main, Germany}



\begin{abstract}
A quantum generalization of the Cooper-Fry recipe is proposed. The
single-particle spectrum arising from relativistic collisions of
particles and nuclei is calculated within the thermal quantum field
theory framework. The starting point of consideration is the
solution of the initial-value problem of particle emission from a
space-like hypersurface. In the following steps, we obtain the
single-particle spectrum using the ``smaller'' Green's function
associated with the fireball medium. Based on this result, several
specific examples of particle emission are considered.
\end{abstract}

\maketitle

\section{Introduction}
\label{sec:introduction}

Models and approaches that are used to describe the processes occurring
in the reaction region during relativistic nucleus collisions are
investigated by comparing the predictions made with experimental data
on one-, two-, and many-particle momentum spectra, which contain information
about the source at an early stage (photons, dileptons) and at the stage
of the so-called {\it freeze-out} (hadron spectra).
The freeze-out hypersurface $\Sigma$ is a kind of spatial surface moving in time,
which is an imaginary boundary between two regions: there is strong dynamics
inside the surface, and outside the surface, particles freely propagate outward.
Wave function at the freeze-out times can be considered as the initial one for its
further history, and, since its further evolution is free, it can be easily
taken into account (we do not discuss final state interactions so far).
Intuitively, free evolution can be reversed, making it possible that the cross
section and other measurable physical quantities are determined through the
initial values of the wave function,
i.e. the values of the wave function at the times of freezing.
Rigorous evaluations give exactly this result.
On the other hand, the strong dynamics acting inside a freeze-out
hyper-surface leads to the creation of a certain quantum state at freeze-out times.
Consequently, the wave function at the freeze-out times is the final state of
strong dynamics.
By representing experimentally measured quantities with the
help of these states, we can study strong interactions in dense and
hot nuclear matter.
Because of this creativity, the separation of the scales of interaction in space
and time, which is carried out using the freeze-out hypersurface, looks
very attractive.

In this article, we consider the emission of hadrons from a space-like
hypersurface, which is related to the initial-value problem.
The particles are emitted by the donor system (fireball), which exhibits
sharp kinetic freezing: space-like and time-like hypersurfaces separate the
excited (interacting) particle system from its noninteracting stage in the
evolutionary process.

\section{Notion of freeze-out hypersurface}
\label{sec:fo-hs}


An important question that can be clarified by studying the reaction zone
is how the space-time boundary of the fireball is
related to the so-called sharp freeze-out hypersurface.
(A discussion of various approaches to the design and use of freeze-out hypersurface
can be found in
\cite{huovinen-2008,adamova-2003,russkikh-ivanov-2006,sollfrank-1999,gersdorff-1986,
strickland-2015,strickland-2016}.)
Usually, the sharp freeze-out hypersurface is defined by the parameter
$P(t,\bs r)$ taking the critical value $P_{\rm c}$ on the hypersurface.
That is, the equation of the hypersurface has a form
\begin{equation}
 P(t,\bs r)\ = \ P_{\rm c}  \,.
\label{eq:anch-begin}
\end{equation}
The parameter to for defining the hypersurface can be selected as:\\
1) The density of particles \cite{adamova-2003}: \quad
$n(t,\bs r)\, =\, n_{\rm c}  \,;$
%
%
%
\\
2) The energy density \cite{russkikh-ivanov-2006,sollfrank-1999}: \quad
$\epsilon(t,\bs r)\, =\, \epsilon_{\rm c}  \,;$
%
%
%
\\
3)  The temperature \cite{huovinen-2008,gersdorff-1986}: \quad
$T(t,\bs r)\, =\, T_{\rm c}  \,.$
%
%
%
\\
It should be noted that any definition of sharp freeze-out is possible precisely
with the selected accuracy.

\begin{figure}[htp]
\begin{center}
\includegraphics[width=1.0\textwidth]{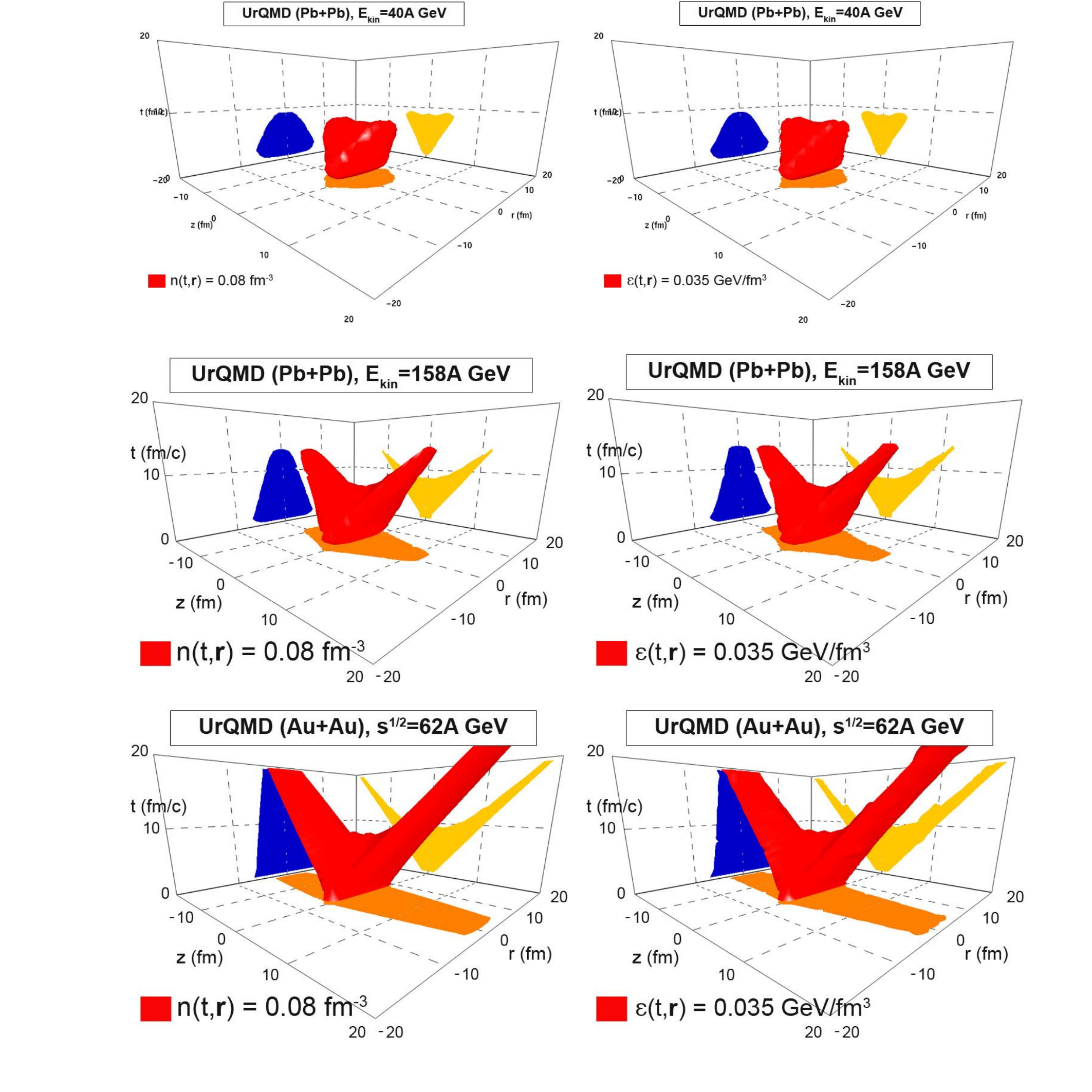}
\end{center}
\caption{
{\it Left panels:}
The hypersurfaces for the critical particle density $n_{\rm c} = 0.08$~fm$^{-3}$.
{\it Right panels:}
The hypersurfaces for the critical energy density
$\varepsilon_{\rm c} = 0.035$~GeV/fm$^3$.
The energies of collision from top row to bottom row, respectively:
$E_{\rm kin} = 40$~AGeV (SPS conditions),
$E_{\rm kin} = 158$~AGeV (SPS conditions),
$\sqrt{s} = 62$~AGeV (RHIC conditions).
Projections to different planes are marked in blue, yellow and light red.
The figures are taken from \cite{anchishkin-2013}.
}
\label{fig:an3}
\end{figure}

Let us list the basic notions of the relativistic kinetic theory \cite{groot-1980}.
%
%
%
%
%
%
Invariant particle number density:
    \begin{equation}
    n(x)\ =\ N^{\mu} (x) \, u_{\mu} (x) \,,
    \label{eq:anch-7}
    \end{equation}
where $N^{\mu}(x) = \int (d^3 p/p^0)  p^{\mu}  f(x,p) =
\left(n_{\rm lab}, n_{\rm lab}\bs v_{_{\rm E}}\right) $ is the particle four-flow
with $f(x,p)$ as the distribution function,
and
$u^{\mu}(x) = N^{\mu}/\left(N^{\nu} N_{\nu}\right)^{1/2}
= \left(\gamma_{_{\rm E}}, \gamma_{_{\rm E}} \bs v_{_{\rm E}}\right)$
is the collective four-velocity (the Eckart definition).
Invariant particle energy density:
    \begin{equation}
    \epsilon(x)\ =\ u_{\mu} (x) \ T^{\mu \nu} (x) \ u_{\nu} (x) \,,
    \label{eq:anch-8}
    \end{equation}
where $T^{\mu \nu}(x) = \int (d^3 p/p^0) p^{\mu} p^{\nu} f(x,p)$ is the
energy-momentum tensor.
With the help of relativistic transport model, the last two equations
(\ref{eq:anch-7}) and (\ref{eq:anch-8}) were used to determine the freeze-out
hypersurface for pions \cite{anchishkin-2013}.
In Fig.~\ref{fig:an3} these freeze-out hypersurfaces, together with the
corresponding projections on three planes, can be seen for different collision energies.
The critical value was taken as $n_{\rm c} = 0.08$~fm$^{-3}$ for the particle density
and $\varepsilon_{\rm c} = 0.035$~GeV/fm$^3$ for the energy density.
The freeze-out hypersurfaces in the figures are represented in
quasi-four-dimensional form, the compactification of spatial transverse coordinates
is made in the form $(x,y) \to r = \sqrt{x^2+y^2}$.
Then the coordinates of the four-dimensional surfaces are $(t,\,r,\,z)$, note the
coordinate $r$ is reflected symmetrically to negative values.

From comparing each pair of plots for the same energy, we can see that
there is no difference in determining the freeze-out hypersurface based
on particle density or based on energy density.
%
This can be considered as a statement that for the pion particle density
$n = n_{\rm c} = 0.035$~GeV/fm$^3$ or the pion energy density
$\varepsilon = \varepsilon_{\rm c} = 0.035$~GeV/fm$^3$
the behavior of a pion system is very close to an ideal gas
(for details see \cite{anchishkin-2013}).

\vspace{5mm}

\section{Radiation of quantum fields from space-like hypersurface: the Cauchy problem}
\label{sec:radiation-qf}

\begin{figure}[htp]
\begin{center}         
\includegraphics[width=0.49\textwidth]{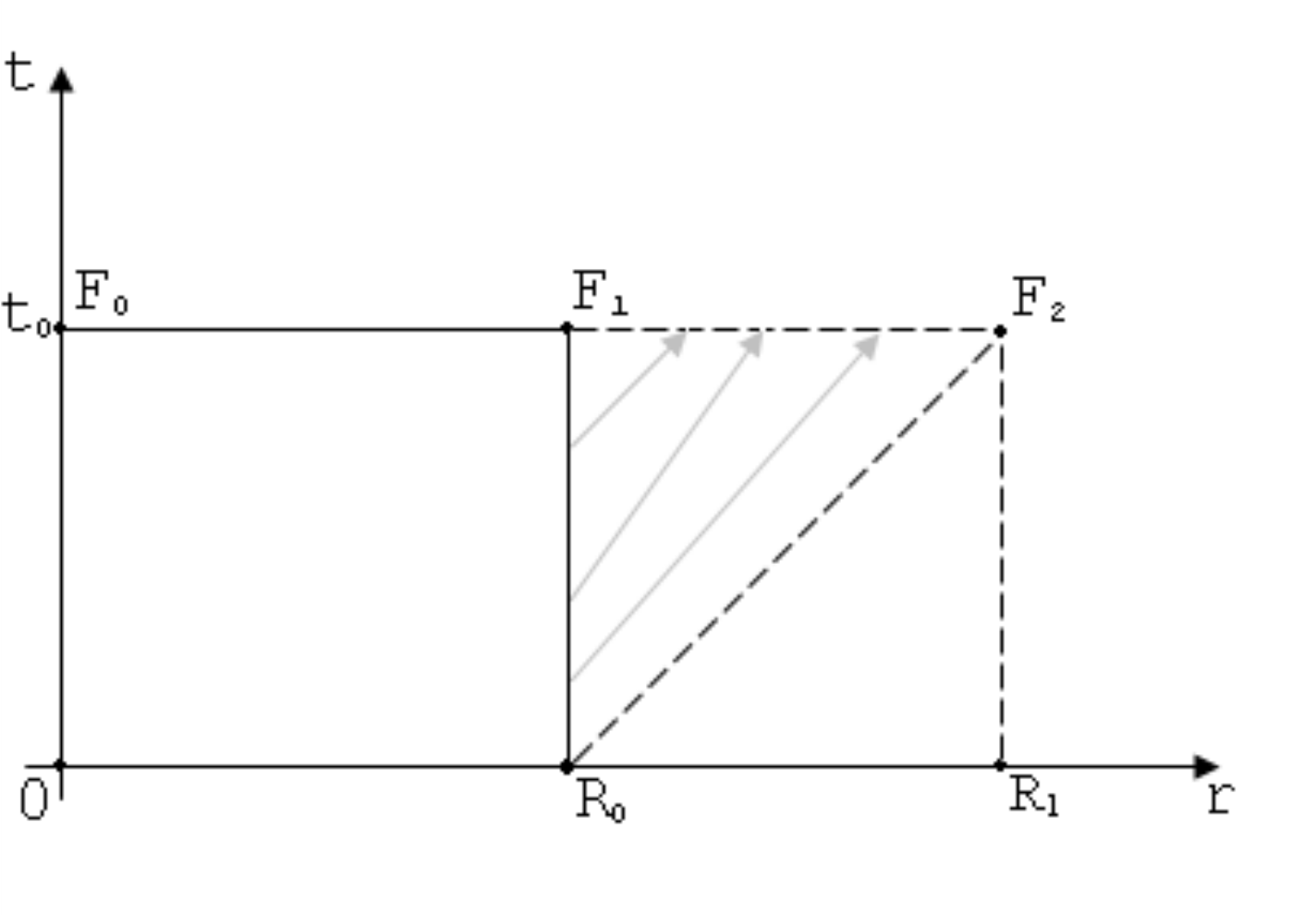}
\includegraphics[width=0.49\textwidth]{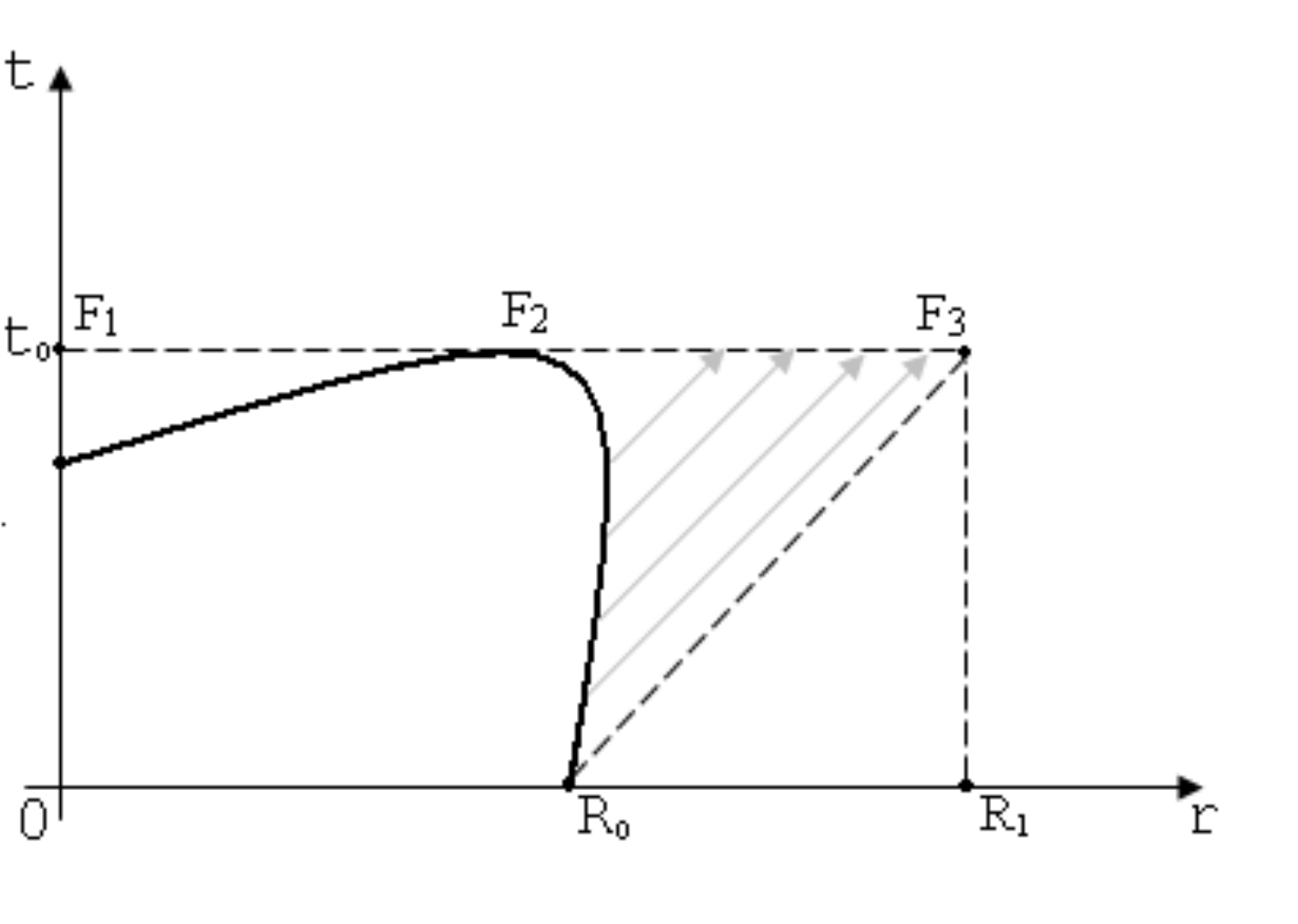}
\end{center}
\caption{ Sketch of a freeze-out hypersurface  for a spherically
symmetric fireball expansion.
{\it Left panel:} Constant initial time $t = t_0$
and constant spatial boundary $R = R_0$.
{\it Right panel:} Dependence of the initial radiation time on the radius
and the dependence of the spatial boundary on time (solid curve).
}
\label{fig:sketch-hs}
\end{figure}

The space-like  hypersurface, together with the time-like hypersurface,
determines the boundary
of the four-dimensional volume occupied by a many-particle system.
Kinetic freeze-out that occurs from a space-like hypersurface (any tangent to
this surface lies outside the light cone) can be treated as an initial-value
problem for the radiation of particles from a fireball.
In the present approach,  we first use a simplified shape of the
freeze-out hypersurface.
The space-like part of the total hypersurface will simply be a surface
of constant time $t=t_0$.
It should be noted that such a condition is a standard approach that is commonly
used in mathematical physics.
On the other hand, we do not lose generality, because the results obtained for an
equal-time hypersurface can be converted to results for a space-like hypersurface
of a given shape, as we will show later.
An example of an equal-time hypersurface for a spherically symmetric expansion of
a fireball is shown in Fig.~\ref{fig:sketch-hs}, on left panel.
One more way is shown, how a complete freeze-out hypersurface, which also includes a
time-like part, can be reduced to an equal-time space-like hypersurface
even in the general case (see the right panel).

So, our goal is to solve the initial-value problem for the propagation of free
particles to the detector when they are ``emitted from the space-like hypersurface''.
In fact, we would like to reproduce and generalize the Cooper-Frye formula
\cite{cooper-frye-PRD-v10-1974} for the emission of particles from a finite
space-time volume.

After freeze-out, the particles propagate without interaction (this is the
freeze-out condition).
This system of ``traveling'' free particles is described by the field operator
$\hat{\varphi} (x)$, which satisfies the Klein-Gordon equation
\begin{equation}
(\partial_\mu \partial^\mu \, +\, m^2) \, \hat{\varphi} (x)\, =\, 0 \,,
\label{3:1i}
\end{equation}
where $\partial_\mu \partial^\mu =\partial_t^2-\vec{\nabla}^2$.
The Cauchy problem or the initial conditions for this equation are specified
on a space-like hyper-surface.
Without loss of generality, for simplicity, this hypersurface has the form
$x^0 = t_0 =$~const (see Fig.~\ref{fig:sketch-hs})
\begin{equation}
\hat{\varphi} (x^0,{\bf x})\Big|_{x^0=t_0}\, =\, \hat{\Phi}_0({\bf x}) \, , \ \ \ \ \ \
\frac{\partial \hat{\varphi}(x^0,{\bf x})}{\partial x^0}\bigg|_{x^0=t_0}\,
=\, \hat{\Phi}_1({\bf x}) \,.
\label{3:2}
\end{equation}
As it is shown in the textbook \cite{vladimirov}, equation (\ref{3:1i}) together
with the initial conditions (\ref{3:2}) can be written  as an
inhomogeneous differential equation (for details see \cite{anch2007})
\begin{equation}
(\partial_\mu \partial^\mu  +m^2) \, \hat{\varphi}(x)\,
=\, \delta(x^0-t_0)\hat{\Phi}_1({\bf x})\,
 +\, \delta'(x^0-t_0)\hat{\Phi}_0({\bf x}) \,,
\label{3:3}
\end{equation}
where using the $\delta'(t)$  function is well known:
$\int dt F(t)\delta'(t)=-\int dt \frac{dF(t)}{dt}\delta(t)$.
This equation can be solved with the help of the retarded Green's
function of the Klein-Gordon equation (see, for example, \cite{schweber}
or \cite{peskin})
\begin{equation}
(\partial_\mu \partial^\mu + m^2)G_R(x-y)\, =\, \delta ^4(x-y) \,,
\label{3:13}
\end{equation}
where
\begin{eqnarray}
G_R(x-y) &=& -\int \frac{d^4k}{(2\pi )^4}\,
\frac{\dis e^{- ik\cdot (x-y)} }{(k_0 + i\delta)^2-{\bf k}^2 - m^2}
\nonumber \\
&=& i\, \theta \left( x^0-y^0 \right) \int \frac{d^3k}{(2\pi )^3 2\omega ({\bf k})} \,
\big[ \,  f_{\bf k}(x) \,f^*_{\bf k}(y)- f^*_{\bf k}(x) \,f_{\bf k}(y)\, \big]
\label{3:12}
\end{eqnarray}
with $f_{\bf k}(x)=e^{- i \omega ({\bf k})x^0+i {\bf k}\cdot {\bf x}}$
and $\omega({\bf k}) = \sqrt{m^2+{\bf k}^2}$.
Hence, for the field operator, $\hat{\varphi}(x)$, from (\ref{3:3})
one obtains
\begin{eqnarray}
\hat{\varphi}(x)
&=& \int d^4y \, G_R(x-y)\, \left[ \delta(y^0-t_0)\hat{\Phi}_1({\bf y})
 + \delta'(y^0-t_0)\hat{\Phi}_0({\bf y}) \right]
\nonumber \\
&=& \int d^4y \, \delta(y^0-t_0) \,
\left[ G_R(x-y)   \frac{\stackrel{\leftrightarrow}{\partial }}{\partial y^0}
       \hat{\Phi}(y^0,{\bf y}) \right] \,,
\label{3:15}
\end{eqnarray}
where we update notations of the initial conditions in the following
way: $\hat{\Phi}_0({\bf y}) = \hat{\Phi}(y^0,{\bf y})\big|_{y^0=t_0}$
and
$\hat{\Phi}_1({\bf y}) = \partial \hat{\Phi}(y^0,{\bf y})/\partial y^0
\big|_{y^0=t_0}$.
Indeed, we associate the field $\hat{\Phi}(y^0,{\bf y})$ with the
interacting system which exists before freeze-out, i.e. at the times
which are less then that on the hyper-surface, which represents the
initial conditions.
In particular case of the flat hyper-surface, $y^0 = t_0$, the field
$\hat{\Phi}(y^0,{\bf y})$ exists at times $y^0 < t_0$.
Hence, this field and its derivatives determine the initial
conditions for particles emitted from the freeze-out hyper-surface
which we described by the field $\hat{\varphi}(x)$.

We note, for the arbitrary space-like hyper-surface, $\sigma(y)$, solution
(\ref{3:15}) looks like \cite{schweber}
\begin{equation}
\hat{\varphi}(x)\, =\, \int_\sigma d\sigma^\mu(y)\, \left[ G_R(x-y)\,
\frac{\stackrel{\leftrightarrow}{\partial }} {\partial y^\mu} \hat{\Phi}(y) \right]\,.
\label{3:15a}
\end{equation}

\bigskip

On the other hand, in the free space
for the system without interaction solution of the Klein-Gordon equation
(\ref{3:1i}) can be written as expansion over eigne functions of the operator of
momentum with the infinite boundary conditions, that coincides with expansion
in the Fourier integral
\begin{eqnarray}
\hat{\varphi} (x^0,{\bf x}) &=& \int \frac{d^3k}{(2\pi)^3\, 2\omega_k}\, \left[b({\bf k})\,
e^{-i\omega_k t+i{\bf k}\cdot {\bf x} }\,
+\, b^+({\bf k})  \, e^{i\omega_k t-i{\bf k}\cdot {\bf x} } \right]
\nonumber \\
&=& \int \frac{d^4k}{(2\pi)^4}\, 2\pi \delta( k^2-m^2)\, e^{-ik\cdot x}\, b({\bf k})\,,
\label{3:5}
\end{eqnarray}
where $\omega_k=\sqrt{m^2+{\bf k}^2}$, \, $b({\bf k})$ and
$b^+({\bf k})$ are the annihilation and creation operators which
satisfy the following commutation relation
\begin{equation}
[b({\bf k}), b^+({\bf p})]\, =\, (2\pi)^3 \, 2\omega_k \, \delta^3({\bf k}-{\bf p})\,.
\label{3:4}
\end{equation}
For the real field, which we consider, these operators should obey
equality
\begin{equation}
b(-{\bf k}) = b^+({\bf k}) \ \ \ \Rightarrow \ \ \
\hat{\varphi}^+ ({\bf x}) = \hat{\varphi} ({\bf x}) \,.
\label{3:6}
\end{equation}
We assume that the detector measures asymptotic momentum eigenstates,
i.e. that it acts by projecting the emitted single-particle state onto
\begin{equation}
\phi^{\rm out}_{\bf k}({\bf x},t)\, =\, e^{ -i\, \omega_k t+i\, {\bf k}\cdot {\bf x}}\,
\equiv\, f_{\bf k}(x) \,,
\label{3:7}
\end{equation}
where $\omega_k=\omega({\bf k})=\sqrt{m^2+{\bf k}^2}$ is the energy of the
particle and we label out-state by the value of measured momentum, i.e. ${\bf k}$.
Evidently, these functions are the solutions of eq.(\ref{3:1i}).
So, we defined the set of the basic functions (asymptotic states)
\begin{equation}
f_{\bf k}(x)\, \equiv\, e^{- i k\cdot x}\,\big| _{k^0=\omega ({\bf k})} \, ,
\label{3:8}
\end{equation}
which are orthogonal with respect to the scalar product
\begin{equation}
\int d^3x\, f^*_{\bf k}(x) \, i \stackrel{\leftrightarrow}{\partial }_{x^0}  f_{\bf p}(x)\,
=\, (2\pi )^3\, 2\omega ({\bf k})\, \delta ^3({\bf k}-{\bf p})\,, \quad
\int d^3x\, f_{\bf k}(x)\, i \stackrel{\leftrightarrow}{\partial }_{x^0} f_{\bf p}(x) = 0\,.
\label{3:9}
\end{equation}
In fact, the expansion (\ref{3:5}) is made in terms of the asymptotic states (\ref{3:8}).
Then, if the field operator, $\hat{\varphi}(x)$, is known the annihilation
and creation operators can be evaluated
\begin{equation}
b({\bf k})\, =\, \int d^3x\, f^*_{\bf k}(x) \, i \stackrel{\leftrightarrow}{\partial }_{x^0}
\hat{\varphi}(x)\,, \quad
b^+({\bf k})\, =\, -\, \int d^3x\, f_{\bf k}(x)\, i \stackrel{\leftrightarrow}{\partial }_{x^0}
\hat{\varphi}^+(x) \,.
\label{3:10}
\end{equation}

The goal of our investigation is a calculation of the single- and
two-particle spectra which can be written with the help of
annihilation and creation operators in the following way
(see, for instance, \cite{GKW,boal,heinz99,anch98,chapman94})
\begin{eqnarray}
P_1({\bf p}) = \left\langle \, b^+({\bf p})\, b({\bf p}) \, \right\rangle\, ,
\quad
P_2({\bf p}_1,{\bf p}_2) = \left\langle \,
b^+({\bf p}_1)\, b^+({\bf p}_2)\, b({\bf p}_2)\, b({\bf p}_1) \, \right\rangle \,,
%
\label{3:16}
\end{eqnarray}
where averaging is taking with respect to the states of the donor
system described by the field $\hat{\Phi}(x)$.
That is why we are interesting in exact expressions of the creation
$b^+({\bf k})$ and annihilation $b({\bf k})$ operators at asymptotic
times, $t \to \infty$.
Inserting solution (\ref{3:15}) into (\ref{3:10}) one obtains
\begin{eqnarray}
b({\bf k}) &=& \lim_{x^0 \to \infty}\, \int d^3x\, d^4y\, \delta(y^0-t_0) f^*_{\bf k}(x)\,
i \frac{\stackrel{\leftrightarrow}{\partial }}{\partial x^0}
\left[ G_R(x-y) \frac{\stackrel{\leftrightarrow}{\partial }}
{\partial y^0}\hat{\Phi}(y) \right]
\nonumber \\
&=& \int d^4y\, \delta(y^0-t_0)\! \left[ f^*_{\bf k}(y)
i \frac{\stackrel{\leftrightarrow}{\partial }}{\partial y^0} \hat{\Phi}(y) \right] ,
\label{3:17}
\end{eqnarray}
and
\begin{eqnarray}
b^+({\bf k}) &=& \lim_{x^0 \to \infty} - \! \int d^3x \, d^4y \, \delta(y^0-t_0) \,
f_{\bf k}(x) i \frac{\stackrel{\leftrightarrow}{\partial }}{\partial x^0}
\left[ G_R(x-y) \frac{\stackrel{\leftrightarrow}{\partial }}
{\partial y^0}\hat{\Phi}^+(y) \right]
\nonumber \\
&=& -\int d^4y \, \delta(y^0-t_0) \! \left[ f_{\bf k}(y)
i \frac{\stackrel{\leftrightarrow}{\partial }}{\partial y^0} \hat{\Phi}^+(y) \right],
\label{3:18}
\end{eqnarray}
where we use representation of the Green's function (\ref{3:12}) and
the orthogonal relations (\ref{3:9}).
It is seen that the projection of the donor field $\hat{\Phi}(y)$ is
taken at the times of the freeze-out and all evolution history of the
free propagation (after freeze-out) is ridded out.

Actually, integration over measure $d^4y \, \delta(y^0-t_0)$ in
(\ref{3:17}) and (\ref{3:18}) is nothing more as integration over
the flat space-like hyper-surface, $y^0=t_0$.
For the arbitrary space-like hyper-surface, $\sigma(x)$, the
annihilation and creation operators expressed through the field
operator $\hat{\Phi}(x)$ look like
\begin{equation}
b({\bf k})\, =\, i\, \int_\sigma d\sigma^\mu(x)\,
\left[ f^*_{\bf k}(x)\, \frac{\stackrel{\leftrightarrow}{\partial }}
  {\partial x^\mu}\, \hat{\Phi}(x) \right] \, ,
\ \ \ \
b^+({\bf k})\, =\, - i\, \int_\sigma d\sigma^\mu(x) \,
\left[ f_{\bf k}(x)\, \frac{\stackrel{\leftrightarrow}{\partial }}
  {\partial x^\mu}\, \hat{\Phi}^+(x) \right] \,.
\label{3:19}
\end{equation}
%

\section{Single-particle spectrum}

With creation and annihilation operators in hand we come to the
evaluation of the single-particle spectrum.
We write
\begin{equation}
2E_p\, \frac{d N}{d^3p} \
=\ \left\langle \, b^+({\bf p}) \, b({\bf p})\, \right\rangle \, ,
\label{eq:23a}
\end{equation}
where $E_p=\omega_p=\sqrt{m^2+\bs p^2}$.
Next, introducing here creation $b^+({\bf k})$ and annihilation $b({\bf k})$
operators from (\ref{3:18}) and (\ref{3:17}) one gets
\begin{eqnarray}
2E_p \frac{d N}{d^3p}
&=&
\int d^4x_1  d^4x_2  \delta(x_1^0-t_0)  \delta(x_2^0-t_0)
\left\langle \left[
f_{\bf p}(x_1) \frac{\stackrel{\leftrightarrow}{\partial }}{\partial x_1^0}
\hat{\Phi}^+ (x_1) \right] \left[  f^*_{\bf p}(x_2)
\frac{\stackrel{\leftrightarrow}{\partial }}{\partial x_2^0}
\hat{\Phi} (x_2) \right] \right\rangle
\nonumber \\
\hspace{-40mm} &=& \int d^4x_1  d^4x_2 \delta(x_1^0-t_0) \delta(x_2^0-t_0)
\bigg\langle \left[
f_{\bf p}(x_1) \frac{\partial }{\partial x_1^0} \hat{\Phi}^+ (x_1)
- \left(\frac{\partial }{\partial x_1^0} f_{\bf p}(x_1)\right) \hat{\Phi}^+ (x_1)\right]
\nonumber \\
&& \hspace{25mm} \times \left[ \,
    f^*_{\bf p}(x_2)\, \frac{\partial }{\partial x_2^0}\, \hat{\Phi} (x_2)
-\, \left( \frac{\partial }{\partial x_2^0}\, f^*_{\bf p}(x_2)\right)\, \hat{\Phi} (x_2)
\right]\, \bigg\rangle \,.
\label{eq:23b}
\end{eqnarray}
All this can be written in a compact form
\begin{equation}
2E_p\, \frac{d N}{d^3p} = \int d^4x_1 \, d^4x_2 \, \delta(x_1^0-t_0) \, \delta(x_2^0-t_0) \,
\left[ \, f_{\bf p}(x_1) \, f^*_{\bf p}(x_2) \,
 \frac{\stackrel{\leftrightarrow}{\partial }}{\partial x_1^0}
 \frac{\stackrel{\leftrightarrow}{\partial }}{\partial x_2^0}
\left\langle \hat{\Phi}^+ (x_1)  \, \hat{\Phi} (x_2) \right\rangle \right] \,.
\label{3:23}
\end{equation}
%
One immediately recognize that angle brackets on the r.h.s. of this
equation give us the correlation function or the lesser Green's function
\cite{kadanoff,chou1985,blaizot}
\begin{equation}
i\, G^<(x_2,x_1) \,=\, \pm \, \left\langle \hat{\Phi}^+(x_1)\, \hat{\Phi}(x_2)
\right\rangle \,,
\label{3:24}
\end{equation}
where the plus sign reads for bosons and the minus sign for fermions (see Appendix).
As an example, in what follows  we consider the Bose statistics.

Hence, for the single-particle spectrum, we obtain the following basic expression
\begin{equation}
2E_p\, \frac{d N}{d^3p}\, =\, i \int d^4x_1\, d^4x_2\, \delta(x_1^0-t_0)\,
\delta(x_2^0-t_0)\,
\left[ \, f_{\bf p}(x_1) \, f^*_{\bf p}(x_2) \,
 \frac{\stackrel{\leftrightarrow}{\partial }}{\partial x_1^0}
 \frac{\stackrel{\leftrightarrow}{\partial }}{\partial x_2^0}\, G^<(x_2,x_1) \right] \,.
\label{3:25}
\end{equation}
As seen from (\ref{3:19}) in covariant form the single-particle spectrum looks like
\begin{equation}
2E_p\, \frac{d N}{d^3p}\, =\, i \int d\sigma^\mu(x_1)\, d\sigma^\nu(x_2)\,
\left[\, f_{\bf p}(x_1)\, f^*_{\bf p}(x_2)\,
 \frac{\stackrel{\leftrightarrow}{\partial }}{\partial x_1^\mu}
 \frac{\stackrel{\leftrightarrow}{\partial }}{\partial x_2^\nu}\, G^<(x_2,x_1) \right] \,.
\label{3:25a}
\end{equation}
The formula (\ref{3:25}) (and (\ref{3:25a})) forms the basis for further
approximations that take into account the special properties of the system
emitting particles.

\subsection*{A note on the description of fermionic radiation}

Let us assume that polarization effects are not taken into account when
registering fermionic particles.
In this case, all polarized states are summed, and the description of the
free state of the propagating particle does not require the details of the
fermionic spin state.
Therefore, for such an experiment, it is quite satisfactory to describe
freely propagating fermions as solutions of the Klein-Gordon equation.

Indeed, let us consider freeze-out of the fermionic system.
After freeze-out the fermions are described by the field
operator $\hat{\psi}(x)$, which satisfies the free Dirac equation
\begin{equation}
(i\gamma^\mu \partial_\mu \,-\, m) \, \hat{\psi}(x) \,=\, 0 \,,
\label{eq:dirac}
\end{equation}
where $\hat \psi$ is the four-spinor (four-component column vector), and the
$\gamma$-matrices satisfy the anticommutation relations
$\{\gamma^\mu,\, \gamma^\nu\} = 2g^{\mu \nu}$ with $g^{\mu \nu}$ as the
metric tensor, sign$(g^{\mu \nu}) = (1,-1,-1,-1)$.
In eq.~(\ref{eq:dirac}) the ``Dirac'' operator can be acted on
from the left by the adjoint operator $(i\gamma^\nu \partial_\nu + m)$.
Using the anticommutation relations of the $\gamma$-matrices, we obtain that
each component of $\hat{\psi}(x)$ must satisfy the Klein-Gordon equation
(\ref{3:1i}), i.e., $(\partial_\mu \partial^\mu + m^2)  \hat{\psi}(x) = 0$.

So, we can conclude that if the experimental registration of fermions does not
fix the polarization of particles, then we can use the solutions of the
Klein-Gordon equation to describe the free propagation of fermions and use
the quantum generalization of the kinetic approach proposed in the present paper
to describe fermion radiation.

\subsection{Single-particle spectrum from homogeneous system}
\label{sec:homogeneous-system}

For a homogeneous system the correlation function can be expanded in the
Fourier integral only with respect to a difference of coordinates
\begin{equation}
G^<(x_2,x_1) \,=\, G^<(x_2-x_1)\, =\, \int \frac{d^4k}{(2\pi)^4}\,
e^{-ik\cdot (x_2-x_1) }\, G^<(k^0,{\bf k}) \,.
\label{3:26}
\end{equation}
%
Substituting then the Green's function in this form into (\ref{3:25})
we obtain
\begin{eqnarray}
2E_p\, \frac{d N}{d^3p} &=&
\int \frac{d^4k}{(2\pi)^4} \, d^3x_1 \, d^3x_2 \,
iG^<(k^0,{\bf k})\, e^{i({\bf p}-{\bf k})\cdot ({\bf x}_1-{\bf x}_2)}\, \times
\nonumber \\
&& \hspace{-15mm} \times \,
\int dx_1^0 \, dx_2^0 \, \delta(x_1^0-t_0) \, \delta(x_2^0-t_0) \,
\left[\, e^{-ip^0 (x_1^0-x_2^0) }\, \frac{\stackrel{\leftrightarrow}{\partial }}
{\partial x_1^0}
 \frac{\stackrel{\leftrightarrow}{\partial }}{\partial x_2^0}\,
 e^{ik^0 (x_1^0-x_2^0)} \right]\,,
\label{3:27}
\end{eqnarray}
where $p^0=\omega_p$.
For two arbitrary functions we have
\begin{equation}
f(t_1,t_2)  \,
 \frac{\stackrel{\leftrightarrow}{\partial }}{\partial t_1}
 \frac{\stackrel{\leftrightarrow}{\partial }}{\partial t_2}\, g(t_1,t_2)
=
 f\, \stackrel{\leftrightarrow}{\partial }_{t_1} \frac{\partial g}
  {\partial t_2}
 -
 \frac{\partial f}{\partial t_2} \, \stackrel{\leftrightarrow}
  {\partial }_{t_1} g
=
 f\, \frac{\partial ^2 g}{\partial t_1  \partial t_2}
 -
 \frac{\partial f}{\partial t_1} \, \frac{\partial g}{\partial t_2}
 -
 \frac{\partial f}{\partial t_2} \, \frac{\partial g}{\partial t_1}
 +
 \frac{\partial ^2 f}{\partial t_1  \partial t_2} \, g
\, .
\label{3:27a}
\end{equation}
Then, expression in square brackets in the last line in (\ref{3:27})
can be easily calculated and one can rewrite (\ref{3:27}) in the following way
\begin{eqnarray}
2E_p\, \frac{d N}{d^3p} &=& \int \frac{d^4k}{(2\pi)^4} d^3x_1\, d^3x_2\,
iG^<(k^0,{\bf k})\, e^{i({\bf p}-{\bf k})\cdot ({\bf x}_1-{\bf x}_2)}
\left( p^0+k^0 \right)^2
\nonumber \\
&& \times \,
\int dx_1^0\, dx_2^0\, \delta(x_1^0-t_0)\, \delta(x_2^0-t_0)\,
e^{-i(p^0-k^0)(x_1^0-x_2^0)} \,,
\label{3:27-2}
\end{eqnarray}
what finally gives
\begin{eqnarray}
2E_p\, \frac{d N}{d^3p}\, =\,
\int \frac{d^4k}{(2\pi)^4}\, i\,G^<(k^0,\bs k)\, \int d^3x_1\, d^3x_2\,
e^{i({\bf p}-{\bf k})\cdot ({\bf x}_1-{\bf x}_2)} \left( E_p + k^0 \right)^2 \,.
\label{3:27t}
\end{eqnarray}
%
%
Expansion (\ref{3:26}) assumes that the donor many-particle system is big enough
to neglect the surface effects.
On the other hand, following the common point of view
\cite{mrowczynski-1990,mrowczynski-1998} we
assume that the Green's function significantly different from zero only when the
differences of arguments, ${\bf x}_1-{\bf x}_2$, are close to zero.
Hence, the integrations on the r.h.s. of (\ref{3:27t}) with respect to
the spatial coordinates ${\bf x}={\bf x}_1-{\bf x}_2$ can be done with
infinite limits, then, it gives the $\delta$-function:
$(2\pi)^3\delta^3({\bf p}-{\bf k})$.
Thus, the single-particle spectrum in a homogeneous system in the rest
frame of the fireball reads
\begin{equation}
2E_p\, \frac{d N}{d^3p}\, =\, V \int \frac{d\omega}{2\pi}\,
\left( E_p + \omega \right)^2\, i\,G^<(\omega,{\bf p}) \,,
\label{3:28}
\end{equation}
where $V =\int d^3X$ with ${\bf X}=({\bf x}_1+{\bf x}_2 )/2$ as the 3D-volume of
the element from which particles are emitted.


To represent the single-particle spectrum (\ref{3:28}) in covariant form,
insert (\ref{3:26}) into (\ref{3:25a}) and get
\begin{equation}
2E_p\, \frac{d N}{d^3p}\, =\, \int \frac{d^4k}{(2\pi)^4} \,  iG^<(k)
\int d\sigma^\mu(x_1)\, d\sigma^\nu(x_2)\, \left[\, f_{\bf p}(x_1)\, f^*_{\bf p}(x_2)\,
 \frac{\stackrel{\leftrightarrow}{\partial }}{\partial x_1^\mu}
 \frac{\stackrel{\leftrightarrow}{\partial }}{\partial x_2^\nu}
 e^{-ik\cdot (x_2-x_1) } \right] \,.
\label{3:28a}
\end{equation}
Calculating the derivatives on the right-hand side of this equation,
we find that the single-particle spectrum of particles emitted by a
homogeneous system in covariant form is written as
\begin{eqnarray}
2E_p\, \frac{d N}{d^3p} \
=&& \hspace{-2mm}
\int \frac{d^4k}{(2\pi)^4}\, (p+k)_\mu (p+k)_\nu \, iG^<(k)
\int_\sigma d\sigma^\mu(x_1)\, d\sigma^\nu(x_2)\, e^{-i(p-k)\cdot (x_1-x_2) }
\nonumber \\
=&& \hspace{-2mm}
\int \frac{d^4k}{(2\pi)^4} \, iG^<(k)
\bigg| \int_\sigma d\sigma^\mu(x) \,  (p+k)_\mu \,e^{-i(p-k)\cdot x }\bigg|^2 \,,
\label{3:28b}
\end{eqnarray}
where $p^0 = E_p = \sqrt{m^2 + \bs p^2}$.

\subsection{ Single-particle spectrum from homogeneous system in
thermodynamic equilibrium }
\label{sec:homogeneous-system-theq}

If the system possesses a global thermodynamic equilibrium, which is
an attribute of the homogeneous system under consideration, the
correlation function $G^<(k)$ can be expressed (see
\cite{mrowczynski-1990,mrowczynski-1998}, in particular in
Ref.~\cite{mrowczynski-1998}, eq.(2.19)) with the help of the
spectral function and the equilibrium distribution function
%
\begin{equation}
i\, G^<(k) \,=\, \theta(k_0)\, A(k)\, f(k^0) \,
-\, \theta(-k_0)\, A(k)\,f(-k^0)
\label{3:29a}
\end{equation}
with
\begin{equation}
f(k^0) \,=\, \frac{1}{e^{ (k^0-\mu)/T} + a} \,,
\label{3:29b}
\end{equation}
where $a = - 1$ for the Bose statistics, $a = 1$ for the Fermi statistics and
$a = 0$ for the classical Boltzmann gas.
To be specific, we consider the Bose statistics below and therefore take
the Bose-Einstein distribution function, that is, $a = - 1$ in (\ref{3:29b}).

If the emitted 3D-area is an element of the freeze-out hyper-surface,
one can assume an absence of interaction on this surface.
For the system of free particles which are in thermal equilibrium
(ideal gas) the spectral function expresses
that all particles are on the mass shell.
Hence, the spectral function of the free fields can be found as
\cite{mrowczynski-1998}
\begin{equation}
A_0(k) \,=\, 2\pi \, \delta(k^2-m^2) \,\big[\, \theta(k^0)-\theta(-k^0)\, \big] \,,
\label{3:30}
\end{equation}
and satisfies the sum-rule \cite{chou1985}
\begin{equation}
\int \frac{dk^0}{2\pi} \, k^0 \,A(k^0,{\bf k}) = 1 \,.
\label{3:30a}
\end{equation}
Using (\ref{3:30}), we can write the representation of the Green's function
$G^<(k)$ given in eq.~(\ref{3:29a}) as follows
\begin{equation}
i\,G^<(k) =
 \frac{\pi}{\omega_k} \, \delta(k_0-\omega_k)\, f_{_{\rm BE}} (k^0) \,
-\, \frac{\pi}{\omega_k} \, \delta(k_0+\omega_k)\, f_{_{\rm BE}} (-k^0) \,.
\label{3:30b}
\end{equation}
For the single-particle spectrum (\ref{3:28}) we need the Green's function,
$G^<(k^0,{\bf p})$, where the measured momentum ${\bf p}$ is put on place of
the momentum ${\bf k}$
(remind, that is due to an integration with infinite limits of the spatial
coordinates ${\bf x}={\bf x}_1-{\bf x}_2$ in (\ref{3:27t}))
\begin{equation}
i\,G^<(\omega,{\bf p}) =
 \frac{\pi}{E_p}\, \delta(\omega - E_p)\, f_{_{\rm BE}}(\omega)\,
-\, \frac{\pi}{E_p}\, \delta(\omega + E_p)\, f_{_{\rm BE}}(-\omega) \,.
\label{3:30c}
\end{equation}
Inserting this in \ref{3:28}) one gets
\begin{equation}
2E_p\, \frac{d N}{d^3p} \,=\, V \int \frac{d\omega}{2\pi}\,
\left( E_p + \omega \right)^2\,
\frac{\pi}{E_p}\,\left[ \delta(\omega - E_p)\, f_{_{\rm BE}}(\omega) \,
-\, \delta(\omega + E_p)\, f_{_{\rm BE}}(-\omega) \right] \,.
\label{3:28-2}
\end{equation}
Due to the presence of the factor $(E_p + \omega_p)^2$ under the integral in
(\ref{3:28-2}), the second term in square brackets does not contribute at all,
because  $\omega = - E_p$ due to the delta-function.
This means that antiparticles do not contribute to the particle spectrum.
Then the single-particle spectrum in the fireball rest frame in the case of
global thermodynamic equilibrium has the form
\begin{equation}
\frac{d N}{d^3p}\ =\  V\,  f_{_{\rm BE}}(E_p) \,,
\label{3:31}
\end{equation}
where $V$ is the spatial volume of the system from which the particles are
emitted, and the system of bosons is taken as an example.

\subsection{The local thermodynamic equilibrium}
\label{sec:loc-equil}

We can go further.
In general case the Fourier expansion of the Green's function looks like
%
\begin{equation}
G^<(x_2,x_1) \,=\, \int \frac{d^4k}{(2\pi)^4}\, e^{-ik\cdot x }\, G^<(X;k) \,,
\label{3:26general}
\end{equation}
where we have made the following transformation of coordinates:
$x = x_1 - x_2$ and $X = (x_1 + x_2)/2$.
Inserting this representation of the Green's function into basic
eq.~(\ref{3:25}) and then calculating the derivatives on the right-hand
side of this equation, we obtain
\begin{eqnarray}
2E_p \frac{d N}{d^3p} = i \! \int \! \frac{d^4k}{(2\pi)^4}
d^4X d^4x \delta(X^0-t_0) \delta(x^0) e^{-i(p-k) \cdot x}  \!
\left[\left( p^0+k^0 \right)^2 G^<(X;k)
+ \frac{\partial^2 G^<(X;k)}{4\, \partial X_0^2} \right] \! ,
\label{3:27-2general-3}
\end{eqnarray}
where $p^0 = E_p = \sqrt{m^2 + \bs p^2}$ and we  used identity
$$
dX^0\, dx^0\, \delta(X^0+x^0/2-t_0)\, \delta(X^0-x^0/2-t_0) \,
=\, dX^0\,  dx^0\, \delta(X^0-t_0)\, \delta(x^0) \,.
$$
In fact, eq.~(\ref{3:27-2general-3}) is a generalization of equation
(\ref{3:27-2}), which was derived for a completely homogeneous system.

We can write also a generalization of the formula (\ref{3:28b}) for
the single-particle spectrum of particles emitted by the
system from a space-like freeze-out hypersurface in covariant form
\begin{eqnarray}
2E_p\, \frac{d N}{d^3p} \
=&& \hspace{-2mm}
i \int \frac{d^4k}{(2\pi)^4}\,
\int_\sigma d\sigma^\mu(x_1)\, d\sigma^\nu(x_2)\, e^{-i(p-k)\cdot (x_1-x_2) }
\nonumber \\
&& \times  \left[(p+k)_\mu (p+k)_\nu \, G^<(X;k)
+ \frac 14 \frac{\partial^2 G^<(X;k)}{\partial X_\mu \partial X^\nu} \right]  \,,
\label{3:28b-2}
\end{eqnarray}
where $p^0 = E_p$ and $X = (x_1 + x_2)/2$.
Here we returned  to the variables of integration $(x_1,x_2)$ and used the
transition $d^4x_1 \delta(x_1^0-t_0) p^0 = d\sigma^\mu(x_1) p_\mu$
and the same for the $x_2$  coordinate.

If the Green's function $G^<(X;k)$ weakly depends on the
sum of arguments, $X$, it can be represented in
the form \cite{mrowczynski-1990,mrowczynski-1998}
\begin{equation}
i\, G^<(X; k) \,=\, \theta(k_0)\, A(X; k)\, f(X;k)
 - \theta(-k_0)\, A(X; k)\, \big[\,f(X;-k) + 1\, \big] \, ,
\label{3:29}
\end{equation}
where $A(X;k)$ is the spectral function and $f(X;k)$ is the
distribution function, which in the kinetic approach is defined as
\begin{equation}
\theta(k_0)\, A(X;k)\, f(X;k) \,=\, \theta(k_0)\, i\, G^<(X;k) \,.
\label{3:29-2}
\end{equation}
The unit in the square brackets on the r.h.s. of (\ref{3:29}) is due
to the tadpole contributions and can be rid out upon subtracting of
the vacuum values as in the case of the vacuum quantum field theory.
Hence, we will skip it in the future.

To take into account the local equilibrium of the radiating system, we use
eq.~(\ref{3:27-2general-3}) with several approximations.
Indeed, in the kinetic approach, the second derivatives of the distribution
function are usually neglected due to the weak dependence on the
coordinates $X = (x_1 + x_2)/2$.
Therefore, we neglect the second derivative of the Green's function on the
right-hand side of (\ref{3:27-2general-3}) and write
\begin{eqnarray}
2E_p \frac{d N}{d^3p} = i \int  \frac{d^4k}{(2\pi)^4} \, d^4X \, \delta(X^0-t_0)\,
(2\pi)^3 \Delta(p-k)\, \left( E_p+k^0 \right)^2 \, G^<(X;k) \,.
\label{eq:27-11}
\end{eqnarray}
Here
\begin{eqnarray}
(2\pi)^3 \, \Delta(p - k)  =  \int d^4x \, \delta(x^0) \, e^{-i(p-k) \cdot x} \,,
\label{eq:ff}
\end{eqnarray}
is a kind of form factor that reflects the finiteness of the spatial volume of
the system.
As was argued before, we assume that the correlation function $G^<(X,x)$ significantly
different from zero only when the differences of arguments, ${\bf x}_1-{\bf x}_2$,
are close to zero.
Therefore, the integrations in (\ref{eq:ff}) can be done with infinite limits, then,
$ \Delta(\bs p - \bs k) \to \delta^3({\bf p}-{\bf k})$.

Further, as before, we assume that on the  freeze-out hypersurface
$G^<(X;k) \approx G_0^<(X;k)$.
Taking into account (\ref{3:30c}) and (\ref{3:29-2}) only for particles
(as we have seen, antiparticles do not contribute) in a system with a slowly varying
inhomogeneity, the Green's function has the form
\begin{equation}
i\,G^<_0(X;k^0,{\bf k}) \,=\,  \frac{\pi}{\omega_k}\, \delta(k^0 - \omega_k)\,
f_{_{\rm BE}}(X;k^0) \,,
\qquad {\rm where}  \qquad  \omega_k = \sqrt{m^2 + \bs k^2} \,.
\label{3:30-11}
\end{equation}
Then, by inserting (\ref{3:30-11}) into eq.~(\ref{eq:27-11}), one can obtain a
formula for calculating the single-particle spectrum in the fireball rest frame
when the radiating system has a weak inhomogeneity
\begin{eqnarray}
2 E_p\, \frac{d N}{d^3p} &=&
 \int  \frac{d^4k}{(2\pi)^4} \, d^4X \, \delta(X^0-t_0)\,
(2\pi)^3 \delta^3({\bf k} - {\bf p})
\left( E_p + k^0 \right)^2 \, \frac{\pi}{\omega_k}\, \delta(k^0 - \omega_k)\,
f_{_{\rm BE}}(X;k^0)
\nonumber \\
&=& \int d^4X\, \delta(X^0-t_0)\, 2 E_p\, f_{_{\rm BE}}(X;E_p)\,.
\label{eq:27-12}
\end{eqnarray}
Obviously, the formula for spectrum (\ref{eq:27-12}) in a homogeneous system
(without dependence on $X$) reduces to (\ref{3:31}).
The right-hand side of eq.~(\ref{eq:27-12}) can be rewritten in covariant notations
%
\begin{equation}
d^4x\, \delta\!\left(x^0-t_0\right)\,p^0\, =\,
\left(d^3x\,p^0 - dtdydz\,p^x - dtdxdz\,p^y - dtdxdy\,p^z\right)
\Big|_{t = t_0 = {\rm const}}\,,
\label{eq:tr-cov}
\end{equation}
where we keep in mind that for the equal-time initial conditions it is valid
$dt = dX^0 = 0$.
So we get
\begin{equation}
d^4X\, \delta\!\left(X^0-t_0\right)\,p^0\, =\, d\sigma_\mu(X)\,p^\mu \,.
\label{eq:tr-cov2}
\end{equation}
By inserting this into eq.~(\ref{eq:27-12}), we rewrite it in covariant
notations as
\begin{eqnarray}
E_p\, \frac{d N}{d^3p}\, =\, \int d\sigma_\mu(X)\,p^\mu\, f_{_{\rm BE}}(X;p\cdot u)\,,
\label{eq:27-14}
\end{eqnarray}
where the particles are on the mass shell $p^0=E_p=\sqrt{m^2+\bs p^2}$,
and $u(X)$ is the four-velocity at the point $X = (X^0, \bs X)$ given on a
space-like hypersurface, on which the initial conditions are now taken into account.
(Remind, to be specific the boson system is considered.)

So, for the particle spectrum we obtain covariant expression (\ref{eq:27-14})
which is valid in any frame.
This expression coincides with the Cooper-Frye formula
\cite{cooper-frye-PRD-v10-1974}.
To obtain it we assume the following:
\\
(1) The many-particle system possess a weak spatially inhomogeneity and
a stationarity of the system is weakly broken.
Due to that reason in the basic equation (\ref{3:27-2general-3})
we have neglected the second derivative of the Green's function
$G^<(X;x)$ with respect to the ``center of mass'' coordinate $X = (x_1 + x_2)/2$.
\\
(2) The many-particle system is at least in the local thermodynamic equilibrium
what is reflected in the dependencies $T(X)$, $\mu(X)$ and $u(X)$.
\\
(3) To obtain eq.~(\ref{eq:27-12}) and then eq.~(\ref{eq:27-14}), we assumed
that the particles are free on the freeze-out hypersurface, which leads to
the approximation $G^<(X;k) \approx G_0^<(X;k)$.
\\
(4) We have assumed that the correlation function $G^<(x_1,x_2)$ differs
significantly from zero only if the differences of the arguments
${\bf x}_1-{\bf x}_2$ are close to zero.
Therefore, integrations in (\ref{eq:ff}) can be performed with infinite limits.
In fact, this is not the case for a system close to a second-order phase
transition, when the correlation length becomes large enough.

Meanwhile, the covariant formula that takes into account all quantum effects is
still (\ref{3:28b-2}).
Even if we take into account approximations (1), (2) and (3)
and therefore use the Green's function in the form
\begin{equation}
i\, G^<(X; k) \,\approx\, i G_0^<(X;k) \,
=\, \theta(k_0)\,  2\pi \, \delta(k^2-m^2)\, f(X;k) \,,
\label{eq:gf-loceq}
\end{equation}
we come to
\begin{eqnarray}
2E_p\, \frac{d N}{d^3p} \
=&& \hspace{-2mm}
\int \frac{d^4k}{(2\pi)^4}\,
\int_\sigma d\sigma^\mu(x_1)\, d\sigma^\nu(x_2)\, e^{-i(p-k)\cdot (x_1-x_2) }
\nonumber \\
&& \times  (p+k)_\mu (p+k)_\nu \, \theta(k_0)\,  2\pi \, \delta(k^2-m^2)\, f(X;k)\,.
\label{3:28b-5}
\end{eqnarray}
After some simplification, we obtain
\begin{eqnarray}
2E_p\, \frac{d N}{d^3p} \,
=\, \int \frac{d^3k}{(2\pi)^3 2\omega_k}\,
\int_\sigma d\sigma^\mu(x_1)\, d\sigma^\nu(x_2)\, e^{-i(p-k)\cdot (x_1-x_2) }
(p+k)_\mu (p+k)_\nu \, f(X;k)\,,
\label{3:28b-6}
\end{eqnarray}
where $p^0 = \sqrt{m^2 + \bs p^2}$ and $k^0 = \omega_k = \sqrt{m^2 + \bs k^2}$.
Reduction of (\ref{3:28b-6}) to the Cooper-Frye formula (\ref{eq:27-14}) is
possible if we can integrate with respect to $\bs x = \bs x_1 - \bs x_2$ on
the right-hand side of eq.~(\ref{3:28b-6}).
This integration will immediately result in delta function
$\delta^3(\bs p - \bs k)$, which helps to convert the integrand to the same form
as in eq.~(\ref{eq:27-14}).
However, this cannot be done for an arbitrary freeze-out hypersurface.
Thus, we can argue that eq.~(\ref{3:28b-6}) is a generalization of the
Cooper-Frye formula (\ref{eq:27-14}), which takes into account quantum effects.
Indeed, one can rewrite eq.~(\ref{3:28b-6}) as
\begin{eqnarray}
2E_p\, \frac{d N}{d^3p} \,
=\, \int_\sigma d\sigma^\mu(x_1)\, d\sigma^\nu(x_2)\, J_{\mu \nu}(x_1 - x_2, \bs p)\,,
\label{3:28b-7}
\end{eqnarray}
were we define the tensor $J_{\mu \nu}$
\begin{eqnarray}
J_{\mu \nu}(x_1 - x_2, \bs p) \,=\, e^{-i p\cdot (x_1-x_2)} \int \frac{d^3k}{(2\pi)^3 2\omega_k}\,
e^{i k\cdot (x_1-x_2)} (p + k)_\mu (p + k)_\nu \, f(X;k)\,.
\label{eq:current}
\end{eqnarray}
As one can see, the integral in (\ref{eq:current}) includes interference of
waves with different momenta $\bs k$, which are projected onto the out-state
with momentum $\bs p$.
There is no such interference in the Cooper-Frye formula (\ref{eq:27-14}),
where $\bs k = \bs p$ is taken.

\subsection{Single-particle spectrum from spatially inhomogeneous
but stationary system}
\label{subsec:spat-inhomog}

For the spatially inhomogeneous, but stationary system, the lesser Green's function
can be represented as $G^<(x_2,x_1) = G^<(x_2^0 - x_1^0;\bs r_1,\bs r_2)$.
Then, after the Fourier transform with respect to the time variable
$x^0 = x_2^0 - x_1^0$
\begin{equation}
G^<(x_2^0 - x_1^0;\bs r_1,\bs r_2)\, =\, \int \frac{d\omega}{2\pi} \,
e^{- i \omega (x_2^0 - x_1^0)} \, G^<(\omega;\bs r_1,\bs r_2) \,,
\label{eq:gf-rel-coord-transform}
\end{equation}
we represent the Fourier transform of the Green's function in the form
\begin{equation}
iG^<(\omega;\bs r_1,\bs r_2)\, =\, A(\omega;\bs r_1,\bs r_2)\,f_{_{\rm BE}}(\omega) \,,
\label{eq:f-transform}
\end{equation}
where $A(\omega;\bs r_1,\bs r_2)$ is the following spectral function
\begin{equation}
A(\omega;\bs r_1,\bs r_2)\, =\, \sum_k \varphi_k(\bs r_1)\,
\varphi_k^*(\bs r_2)\, \frac{\pi}{\epsilon_k} \delta(\omega - \epsilon_k) \,.
\label{eq:sf}
\end{equation}
Here $\{\varphi_k(\bs r)\}$ is the orthogonal and complete set of functions
describing the spatial distribution of particles in a fireball.

Let us check the reduction to the homogeneous case.
When particles on the freeze-out hypersurface are free, their wave
functions will be plane waves
$\varphi_k(\bs r) \to (1/\sqrt{V})\exp{(i \bs k \cdot \bs r)}$ and
hence $\epsilon_k \to E_k = \sqrt{m^2 + \bs k^2}$.
By inserting this into eq.~(\ref{eq:sf}) one gets
\begin{equation}
A(\omega;\bs r_1,\bs r_2) \,=\,  \int \frac{d^3k}{(2\pi)^3} \,
e^{i \bs k \cdot (\bs r_1 - \bs r_2)} \, \frac{\pi}{E_k} \, \delta(\omega - E_k) \,.
\label{eq:checking}
\end{equation}
Making the Fourier transform with respect to the relative coordinate
$\bs r = \bs r_1 - \bs r_2$ we come to the spectral function
\begin{eqnarray}
A(\omega,\bs p) \,=\,  \int \frac{d^3k}{(2\pi)^3} \, d^3 r \,
e^{-i (\bs p - \bs k) \cdot \bs r}\, \frac{\pi}{E_k} \, \delta(\omega - E_k)
=\, \frac{\pi}{E_p} \, \delta(\omega - E_p) \,.
\label{eq:checking2}
\end{eqnarray}
That is, in the case of a homogeneous system in thermal equilibrium, we arrive
at the same free spectral function as in (\ref{3:30}).

To calculate the spectrum of particles we use the basic eq.~(\ref{3:25}), where
we introduced the Green's function in the form (\ref{eq:gf-rel-coord-transform})
and using representation (\ref{eq:f-transform}) we obtain
\begin{eqnarray}
2E_p\, \frac{d N}{d^3p} &=& \int d^4x_1 \, d^4x_2 \,
\delta(t_1-t_0) \, \delta(t_2-t_0) \, \int_{-\infty}^\infty \,
\frac{d\omega}{2\pi}\ A(\omega;\bs r_1,\bs r_2)\, f_{_{\rm
BE}}(\omega)\, \times
\nonumber \\
&& \times\, \left[ \, f_{\bf p}(t_1,\bs r_1) \, f^*_{\bf p}(t_2,\bs r_2) \,
 \frac{\stackrel{\leftrightarrow}{\partial }}{\partial t_1}
 \frac{\stackrel{\leftrightarrow}{\partial }}{\partial t_2} \,
 e^{-i\omega (t_2-t_1)} \right] \,.
\label{eq:spectrum1}
\end{eqnarray}
%
We calculate two-side derivatives (remind,
$f_{\bf p}(x) = \exp{(- i E_p x^0 + i \bs p \cdot \bs x)}$)
\[
\left[ \, e^{-iE_p (t_1-t_2)} \,
 \frac{\stackrel{\leftrightarrow}{\partial }}{\partial t_1}
 \frac{\stackrel{\leftrightarrow}{\partial }}{\partial t_2} \,
 e^{-i\omega (t_2-t_1)} \right]\,
 =\, (\omega + E_p)^2\, e^{-i(E_p - \omega) (t_1-t_2)} \,.
\]
%
Then we put everything together and rewrite (\ref{eq:spectrum1}) as
\begin{equation}
2E_p\, \frac{d N}{d^3p} = \int_{-\infty}^\infty \,
\frac{d\omega}{2\pi}\ f_{_{\rm BE}}(\omega)\, (\omega + E_p)^2\,
\int d^3r_1 \, d^3r_2 \, e^{i\bs p \cdot (\bs r_1-\bs r_2)}
A(\omega;\bs r_1,\bs r_2) \,.
\label{eq:spectrum2}
\end{equation}
Using the spectral function representation (\ref{eq:sf}), we can go further
\begin{eqnarray}
2E_p\, \frac{d N}{d^3p} &=& \int_{-\infty}^\infty \, \frac{d\omega}{2\pi} \,
f_{_{\rm BE}}(\omega)\, (\omega + E_p)^2\, \sum_k \varphi_k(\bs
p)\, \varphi_k^*(\bs p)\, \frac{\pi}{\epsilon_k} \delta(\omega -\epsilon_k)
\nonumber \\
&=& \sum_k\, \big|\, \varphi_k(\bs p)\, \big|^2\, \frac{1}{2 \epsilon_k} \,
(\epsilon_k + E_p)^2\, f_{_{\rm BE}}(\epsilon_k)\,,
\label{eq:spectrum3}
\end{eqnarray}
where
\begin{equation}
\varphi_k(\bs p)\, =\, \int d^3r \,e^{i\bs p \cdot \bs r}\, \varphi_k(\bs r)\,.
\label{eq:ft2}
\end{equation}
If the quasi-momentum $\bs k$ is assumed to be continuous variable, then the
single-particle spectrum has the form
\begin{equation}
2E_p\, \frac{d N}{d^3p}\, =\, V \int \frac{d^3k}{(2\pi)^3}\,
\big|\, \varphi_k(\bs p)\, \big|^2\, \frac{1}{2 \epsilon_k} \,
(\epsilon_k + E_p)^2\, f_{_{\rm BE}}(\epsilon_k)\,.
\label{eq:spectrum4}
\end{equation}

For example, for a system in a box $V = L^3$ with the Dirichlet
boundary conditions, the functions $\varphi_n(\bs r)$ satisfy the
stationary Klein-Gordon equation
\begin{equation}
\left( \epsilon_{\bs n}^2 - m^2 + \bs \nabla^2 \right) \varphi_{\bs n}(\bs r)  \,=\, 0 \,.
\label{eq:k-g}
\end{equation}
The solutions of this equation normalized to unity have the form
\begin{equation}
\varphi_{\bs n}(\bs r)  \,=\, \sqrt{\frac{8}{V}} \,
\sin\left(k_x \,x\right) \,
\sin\left(k_y \,y\right) \,
\sin\left(k_z \,z\right) \,,
\label{eq:solution-kg-box}
\end{equation}
where $\bs n = (n_x,n_y,n_z)$ with $n_i = 0, 1, 2, \ldots$ and
$\bs k = \left(\frac{\pi}{L} n_x, \frac{\pi}{L} n_y, \frac{\pi}{L} n_z\right)$.
Each quantum state of the system, marked with the symbol $\bs n$, has a self energy
$\epsilon_{\bs n} = \sqrt{m^2 + \left(\frac{\pi}{L}\right)^2 \bs n^2}$.
Then, using the set of wave functions $\varphi_{\bs n}(\bs r)$ and the set of
self-energies $\epsilon_{\bs n}$, one can calculate the spectrum
(\ref{eq:spectrum3}) or for large volume of the box can use eq.~(\ref{eq:spectrum4}).

For greater realism, one can consider a cylinder extended along the $z$ axis,
or a cylinder with an elliptical cross section to take into account the
azimuthal dependence of particle radiation.

\subsection{Radiation of particles from a system with a finite lifetime}
\label{subsec:fin-lt}

To reflect the finite lifetime of the system, we can use the parameterization
of the spectral function in the spirit of the Bright-Wigner distribution
function since each particle of the system is characterized by the smearing
of its energy in the same way as resonances.
It can be done by smearing the delta function in eq.~(\ref{3:30})
\begin{eqnarray}
\delta(k_0^2-\omega_k^2)\, \to \, A_\gamma(k) \,=\,
\frac{1}{\pi} \, \frac{m\gamma}{(k_0^2-\omega_k^2)^2 + (m\gamma)^2} \,,
\label{3:101}
\end{eqnarray}
where $\gamma=1/\tau$ with $\tau$ as the lifetime of the donor
many-particle system.
Here we use the limit equality
$\delta(x) = \lim_{\gamma \to 0} \frac 1\pi \frac{\gamma}{x^2 + \gamma^2}$.
 %
 %
%
%
Then the spectral function of free fields (\ref{3:30}) is transformed
into a new spectral function $A(k)$, which reflects the finite life time $\tau$
of the system
\begin{equation}
A_0(k) \, \to \, A(k) \,=\, 2\pi \,  A_\gamma(k)\, \left[\theta(k^0) \,-\,
\theta(-k^0) \right] \,.
\label{3:104}
\end{equation}
Let us check the sum rule (\ref{3:30a}) for the resulting parametrization
(\ref{3:104}) of the spectral function
\begin{eqnarray}
\int_{-\infty}^\infty \frac{dk^0}{2\pi} \, k^0 \,A(k^0,{\bf k})
&=&
\int_0^\infty \frac{dk^0}{\pi} \,
\frac{k_0\, m\gamma}{(k_0^2-\omega_k^2 )^2+(m\gamma)^2} \,
-\, \int_{-\infty}^0 \frac{dk^0}{\pi} \,
\frac{k_0\, m\gamma}{(k_0^2-\omega_k^2 )^2+(m\gamma)^2}
\nonumber \\
&=&
2\int_0^\infty \frac{dk^0}{\pi} \,
\frac{k_0\, m\gamma}{(k_0^2-\omega_k^2 )^2+(m\gamma)^2}
\nonumber \\
&=&
\int_0^\infty \frac{ds}{\pi} \, \frac{ m\gamma}{(s-\omega_k^2 )^2+(m\gamma)^2}\,
=\, 0.5 + \frac 1\pi \arctan\left( \frac{m \gamma}{\omega_k^2} \right)
\approx 1 \,,
\label{3:105}
\end{eqnarray}
%
where the last approximation is valid if the inequality
\begin{equation}
m\gamma \, \ll \, \omega_k^2 \, , \ \ \ \to \ \ \
\gamma \, \ll \, m+\frac{{\bf k}^2}{m} \,.
\label{3:106}
\end{equation}
Since the inverse lifetime of the system $\gamma$ is an average
value, i.e. we do not separate $\gamma$ and $\langle\gamma\rangle$,
the inequality (\ref{3:106}) can be considered as averaged for a thermal
many-particle system at a temperature $T$:
\begin{equation}
\langle\gamma\rangle \, \ll \, m + 2\left\langle\frac{{\bf k}^2}{2m}\right\rangle
= m+3T \,,
\label{3:107}
\end{equation}
%
where the last equality uses the nonrelativistic relation
$\big\langle\frac{{\bf k}^2}{2m}\big\rangle=\frac32 T$.
For example, for the thermal system of pions $m_\pi=140$~MeV, which
is created in collisions of relativistic nuclei, the freeze-out temperature
$T\simeq 150$~MeV is known.
Therefore, we can make the following estimate of the allowable width $\gamma$:
$1/\tau = \gamma \, \ll \, 590$~MeV.
With good accuracy ($\int \frac{dk^0}{2\pi} k^0 A(k) \approx 0.95$),
the lower limit of the lifetime can be taken equal to
$\tau_c\simeq 2$~fm/c ($\gamma_c\simeq 100$~MeV).
Then for the total lifetime of the multipion system $\tau$ greater than $\tau_c$ the
spectral function (\ref{3:104}) satisfies the sum rule (\ref{3:30a}).

Following the spirit of the Breit-Wigner distribution function, we only
use it for positive energies.
With this in mind, we write the Green's function
%
\begin{eqnarray}
i\,G^<(k_0,\bs k) \,=\, 2\pi \theta(k_0)  \, A_\gamma(k_0,\bs k)\, f_{_{\rm BE}}(k_0)\,.
\label{3:108}
\end{eqnarray}
Then the single-particle spectrum (\ref{3:28}) can be expressed as
(note, we introduce into (\ref{3:28}) the Green's function in the form
$G^<(k_0,\bs p)$)
\begin{eqnarray}
2E_p \, \frac{d N}{d^3p} \,=\, V \int_0^\infty dk^0\, \left( E_p+k_0 \right)^2\,
A_\gamma(k_0,\bs p)\, f_{_{\rm BE}}(k_0) \,.
\label{3:109}
\end{eqnarray}
And, finally, we get the spectrum of particles emitted from the system with
a finite lifetime, $\tau = 1/\gamma$,
\begin{equation}
2E_p \, \frac{d N}{d^3p}\,
=\, V \int_0^\infty \frac{dk^0}{\pi}\, \left( E_p+k_0 \right)^2\,
\frac{ m\gamma }{(k_0^2-E_p^2)^2+(m\gamma )^2}\, f_{_{\rm BE}}(k^0)\,.
\label{3:109a}
\end{equation}

Let us check the convergence to the previous result.
When $\gamma \to 0$, the integrand in (\ref{3:109a}) is converted as
\begin{eqnarray}
2E_p \, \frac{d N}{d^3p}
&=& V \int_0^\infty dk^0 \, \left( E_p+k_0 \right)^2\,
\delta(k_0^2-E_p^2)\, f_{_{\rm BE}}(k_0) \,.
\label{3:109-d}
\end{eqnarray}
That leads to
\begin{eqnarray}
\frac{d N}{d^3p} \,=\, V\,  f_{_{\rm BE}}(E_p)\,.
\label{3:109-2d}
\end{eqnarray}
This formula coincides with (\ref{3:31}), which gives the single-particle
spectrum in the fireball rest frame in the case of global thermodynamic
equilibrium.

\section{Applications}
\label{subsec:applications}

The goal of this section is to compare on the basis of the specific
examples, the Cooper-Frye formula (\ref{3:31}) and a quantum
generalization of the kinetic approach given in eq.~(\ref{3:28b-6}).

\subsection{Constant time hypersurface: Rapidity distribution}
\label{subsec:fin-lt}

In case, of the constant-time space-like hypersurface, i.e., $x_0 =
t_0 =$~const on the hypersurface, we have $d\sigma^\mu(x) = d^4x
\delta(x_0 - t_0)$.
To make the comparison as transparent as possible, we consider the pion
system in the global thermodynamic equilibrium.
In this case formula (\ref{3:28b-6}) looks like
\begin{eqnarray}
2E_p\, \frac{d N}{d^3p} &=& \int \frac{d^3k}{(2\pi)^3 2\omega_k}\,
f_{_{\rm BE}}(\omega_k)
 \left| \int_\sigma d\sigma^\mu(x)\, (p+k)_\mu \, e^{-i(p-k)\cdot x } \right|^2
 \nonumber \\
&=& \int \frac{d^3k}{(2\pi)^3 2\omega_k}\, f_{_{\rm BE}}(\omega_k)\,
(E_p + \omega_k)^2
 \left| \int d^2x_T\,  \, e^{i(\bs p_T - \bs k_T)\cdot \bs x_T }
 \int_0^{L_z} dz\, e^{i(p_z - k_z)z }\right|^2  \,,
\label{app:t0-const}
\end{eqnarray}
where $p^0 = E_p =\sqrt{m^2 + \bs p^2}$ and $k^0 = \omega_k =
\sqrt{m^2 + \bs k^2}$.
Our goal is to calculate rapidity
distribution, then let us assume that in the transverse direction, a
2D volume is big enough that the integration over the transverse
spatial coordinates in (\ref{app:t0-const}) leads to the
delta-function. In detail, it looks like
\begin{equation}
\left|\int d^2x_T\,  \, e^{i(\bs p_T - \bs k_T)\cdot \bs x_T }
\right|^2 \, =\, (2\pi)^2 \delta^2(\bs p_T - \bs k_T) \int_{(S_T)}
d^2x_T\, e^{i(\bs p_T - \bs k_T)\cdot \bs x_T } \, =\, S_T \,
(2\pi)^2 \delta^2(\bs p_T - \bs k_T)  \,, \label{app:int-transv}
\end{equation}
where $S_T$ is the square of the transverse cross section. After
inserting the obtained $\delta$-function into
eq.~(\ref{app:t0-const}) we get $k_T = p_T$ and consequently
$m_T(k_T) = m_T(p_T)$.
Integration over $z$-direction over the
longitudinal size of the system $L_z$ gives a one-dimensional form
factor
\begin{equation}
\int_0^{L_z} dz\, e^{i(p_z - k_z)z } \, =\, \frac{e^{i (p_z -
k_z)L_z} - 1}{i(p_z - k_z)} \, =\, e^{i(p_z - k_z)L_z/2} \
\frac{2\sin{\left[ (p_z - k_z)L_z/2\right]}}{i(p_z - k_z)}  \,.
\label{app:ff}
\end{equation}
In fact, this is a particular example of the relativistic form
factor introduced in eq.~(\ref{eq:ff}).

\begin{figure}[t]
\centering
\includegraphics[width=0.31\textwidth]{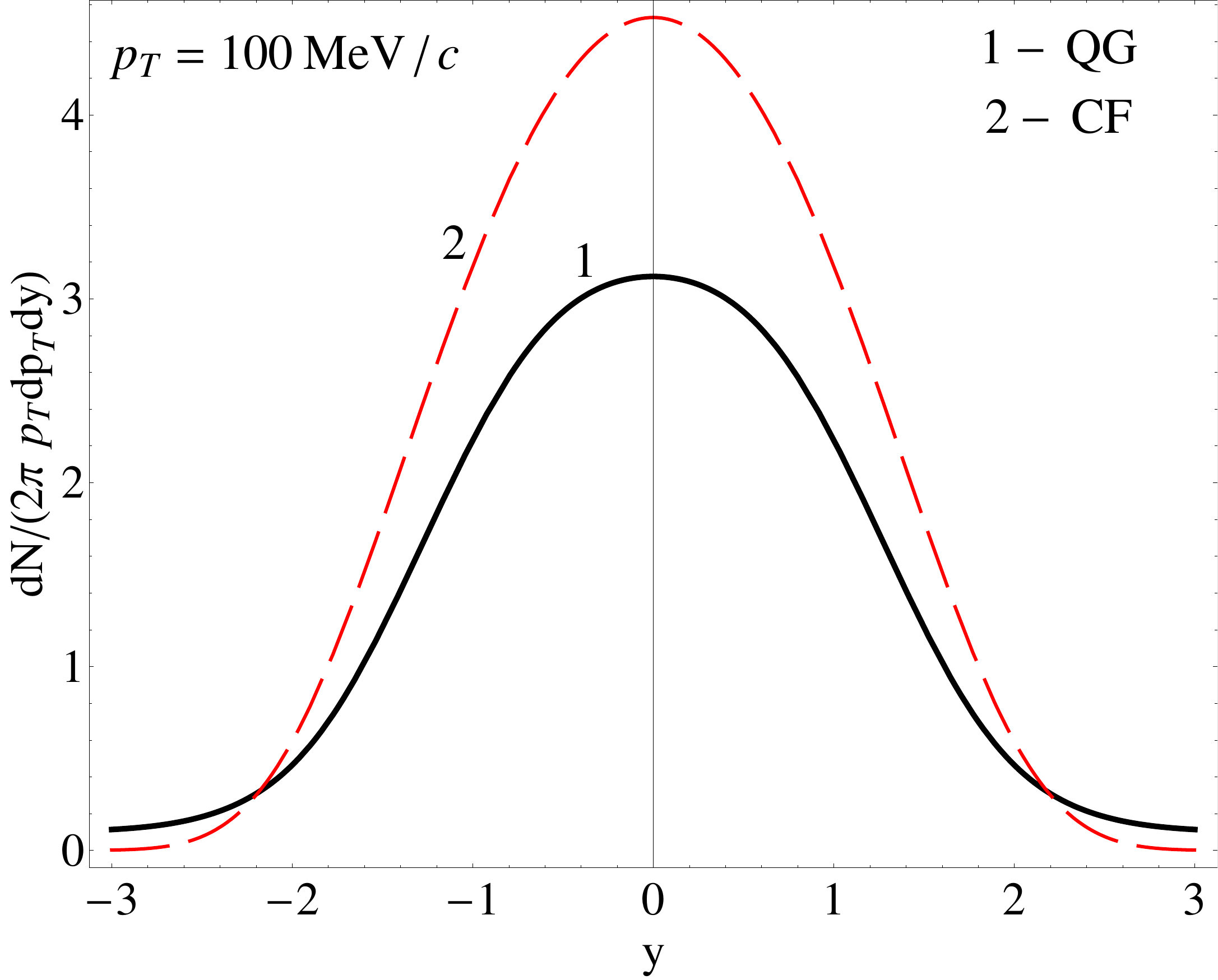}
\includegraphics[width=0.31\textwidth]{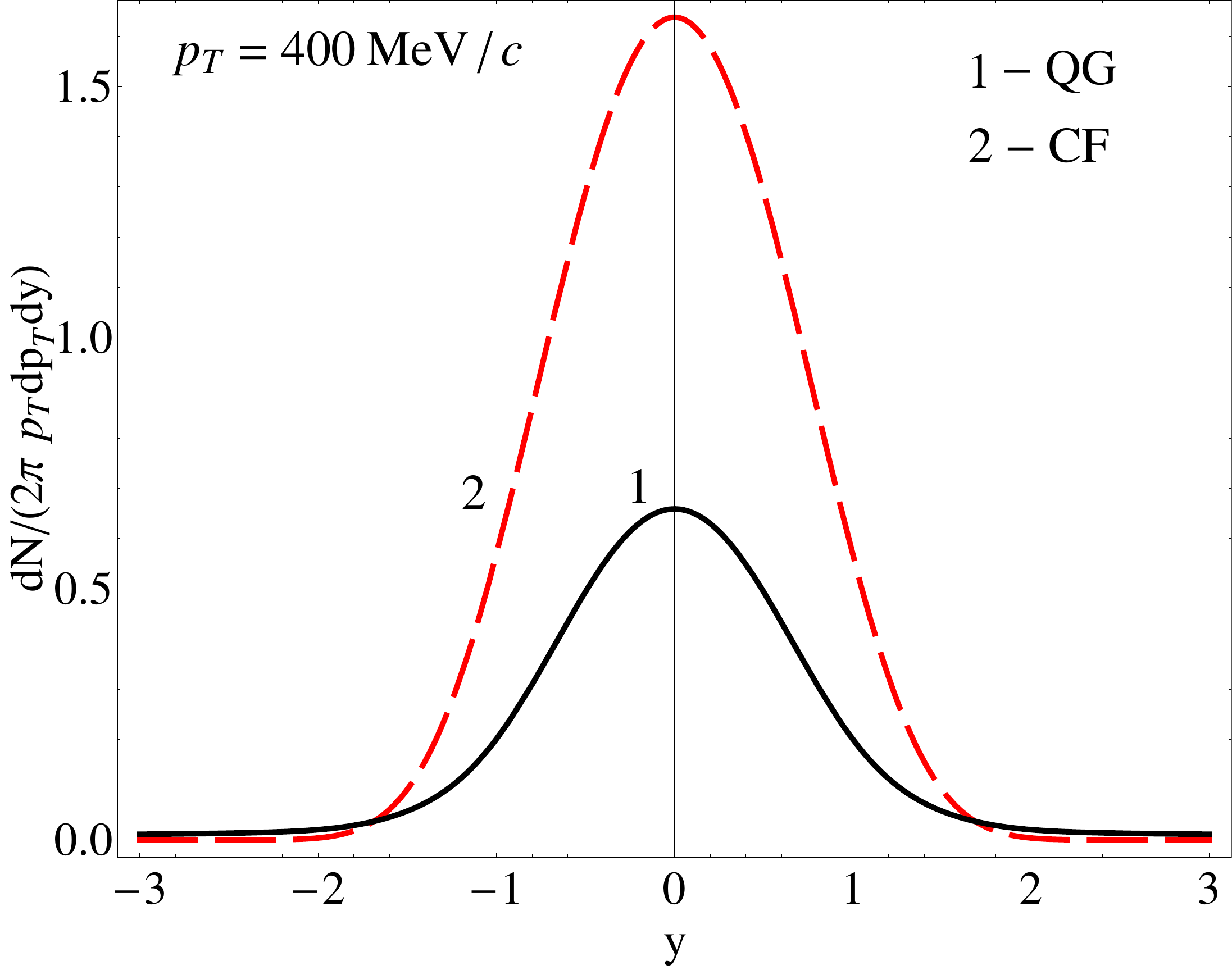}
\includegraphics[width=0.31\textwidth]{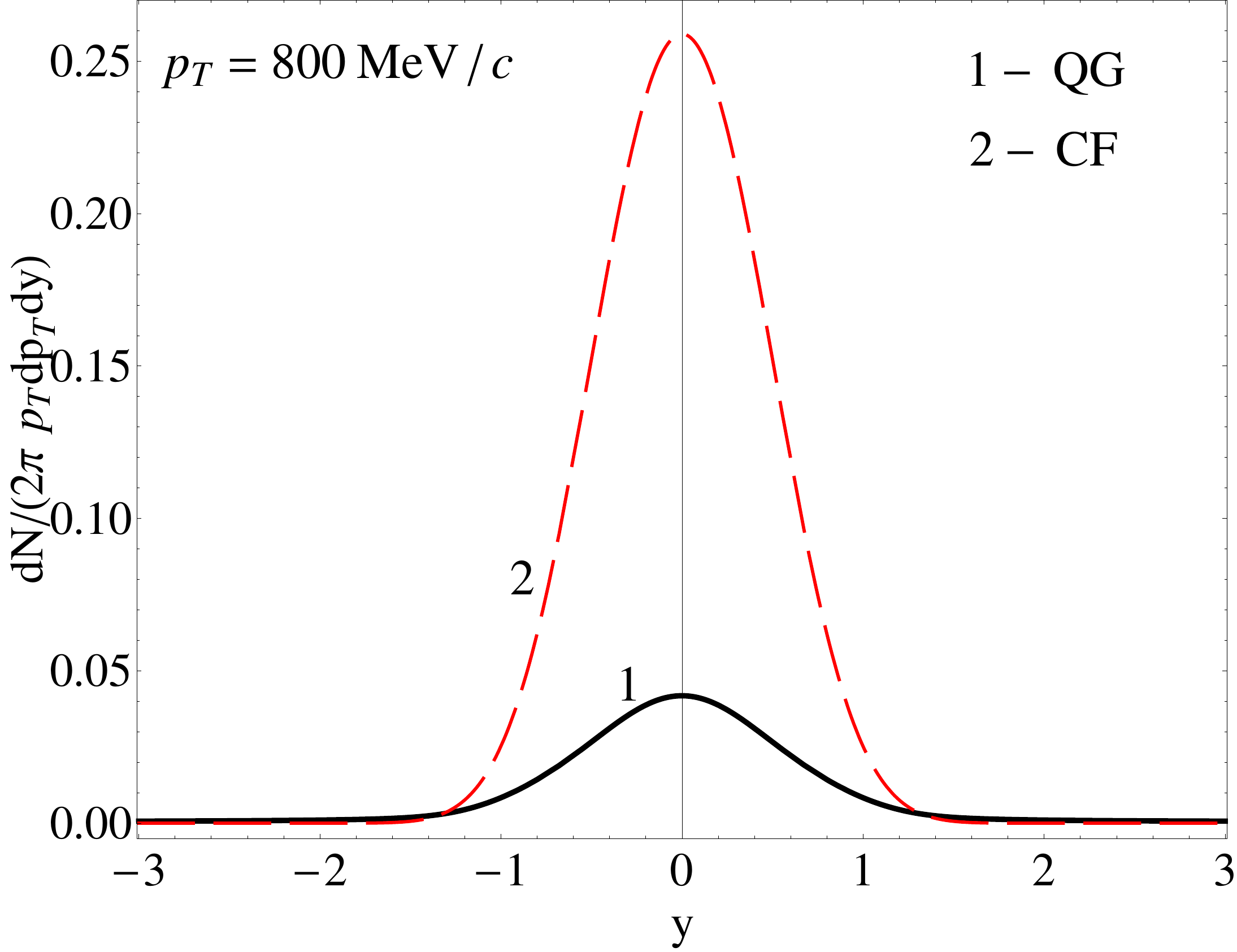}
\caption{ Rapidity distribution of pions within the Cooper-Frye
approach (red dashed curves, marked as CF) and within the quantum
generalization of the kinetic approach (solid black curves, marked
as QG).
Every panel corresponds to a fixed transverse momentum of
the pion, $p_T = 100,\, 400,\, 800$~MeV/c, from the left to right,
respectively.
The longitudinal size of the system $L_z = 10$~fm, and
$T = 160$~MeV, $\mu = 0$. } \label{fig:y-distrib}
\end{figure}

We are going to consider the rapidity distributions and transverse
momentum spectra.
That is why it is reasonable to make
transform of the momentum coordinates: $(p_x,p_y,p_z) \to (\varphi,
p_T, y)$, where $\varphi$ is the azimuthal angle, $p_T = \sqrt{p_x^2
+ p_y^2}$ is the transverse momentum and $\tanh{y} = p_z/E_p$ is the
rapidity.
For the differential, we get $d^3p = 2\pi E_p p_T dp_T
dy$, where we integrate over the azimuthal angle assuming the
azimuthal symmetry.
Then, inserting (\ref{app:int-transv}) and
(\ref{app:ff}) into eq.~(\ref{app:t0-const}) for dependence of the
spectrum on variables $(p_T,y)$ we get
\begin{eqnarray}
\frac{d N}{2\pi  p_T dp_T dy} \,=\, S_T \int \frac{dy_k}{2\pi} \,
f_{_{\rm BE}}\big(m_T(p_T) \cosh{y_k}\big)\,
\big(\cosh{y} + \cosh{y_k}\big)^2 \, F_L(y_k;y,p_T) \,,
\label{app:t0-const3}
\end{eqnarray}
where
\begin{eqnarray}
F_L(y_k;y,p_T) \,=\, \left[ \frac{\sin{\left[ L_z m_T(p_T) (\sinh{y}
- \sinh{y_k})/2\right]}} {(\sinh{y} - \sinh{y_k})} \right]^2 \,,
\label{app:longit-ff}
\end{eqnarray}
is the longitudinal form factor and $\tanh{y_k} = k_z/\omega_k$ is
the rapidity of particle with the momentum $\bs k$.
To derive eq.~(\ref{app:t0-const3}) we have used that $m_T(k_T) = m_T(p_T)$,
what results in $k_z =  m_T(p_T) \sinh{y_k}$, and then $dk_z =
m_T(p_T) \cosh{y_k} \, dy_k$.

On the other hand, the Cooper-Frye spectrum in the fireball rest
frame in the case of global thermodynamic equilibrium has the form
(\ref{3:31}).
In the Cooper-Frye approach, the rapidity distribution
for the fixed transverse momentum looks like
\begin{eqnarray}
\frac{d N}{2\pi p_T dp_T dy} \,=\, S_T \, L_z\, m_T(p_T) \cosh{y} \,
f_{_{\rm BE}}\big(m_T(p_T) \cosh{y}\big) \,. \label{app:cf-pt}
\end{eqnarray}
%
Here, we have factorized the volume of the system $V$ into the
transverse cross section $S_T$ and the longitudinal size of the
system $L_z$ that is $V = S_T L_z$.
To compare the spectrum (\ref{app:t0-const3}) with the spectrum
(\ref{app:cf-pt}) we set $S_T = 1$.
At the same time, to keep right dimensionality in the
adopted system of units (c = $\hbar = 1$) we take the numerical
value of $L_z$ as: $N[L_z]$~fm~$\to (N[L_z]/C_c)$~MeV$^{-1}$, where
$C_c = 197.3\,$ (remind, $C_c\cdot$MeV$\cdot$fm = 1).

The calculated rapidity distributions are depicted in
Fig.~\ref{fig:y-distrib} in three panel in a correspondence with
three values of the fixed transverse momentum of the $\pi$-meson,
$p_T = 100,\, 400,\, 800$~MeV/c, respectively, and for a given
longitudinal size of the system $L_z = 10$~fm.
Parameters of the Bose-Einstein distribution function are
$T = 160$~MeV, $\mu = 0$.
Black solid curves, which are marked as QG, reflect calculations
with the help of eq.~(\ref{app:t0-const3}), i.e., it is a quantum
generalization of the kinetic approach.
Red dashed curves, which are marked as CF, reflect calculations with
the help of eq.~(\ref{app:cf-pt}).
It is seen that the difference between the two distributions goes up
with an increase in the transverse momentum.

\subsection{Radiation of particles from a system with a finite lifetime}
\label{subsec:fin-lt}

\begin{figure}[t]
\centering
\includegraphics[width=0.44\textwidth]{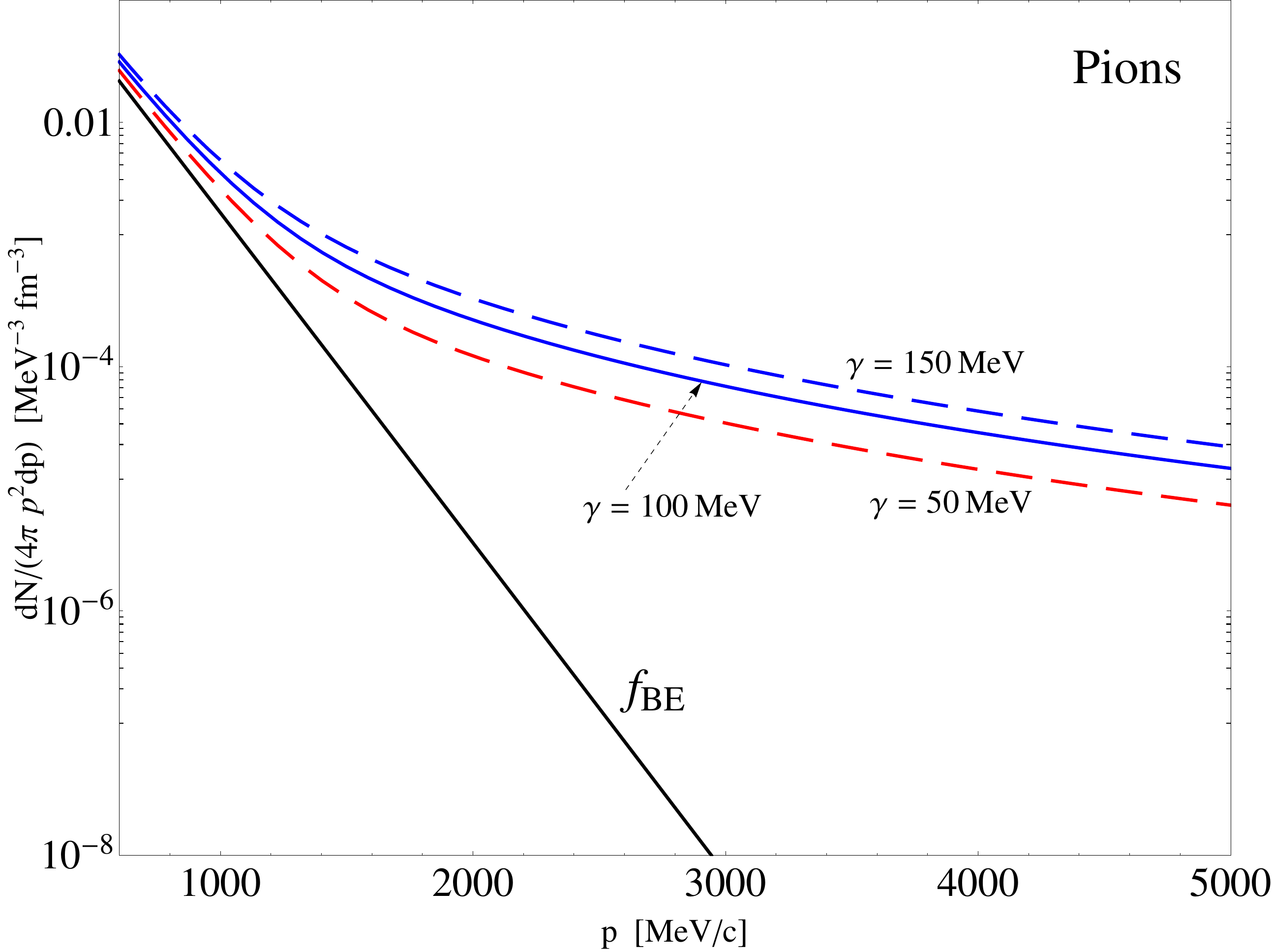}
\hspace{2mm}
\includegraphics[width=0.46\textwidth]{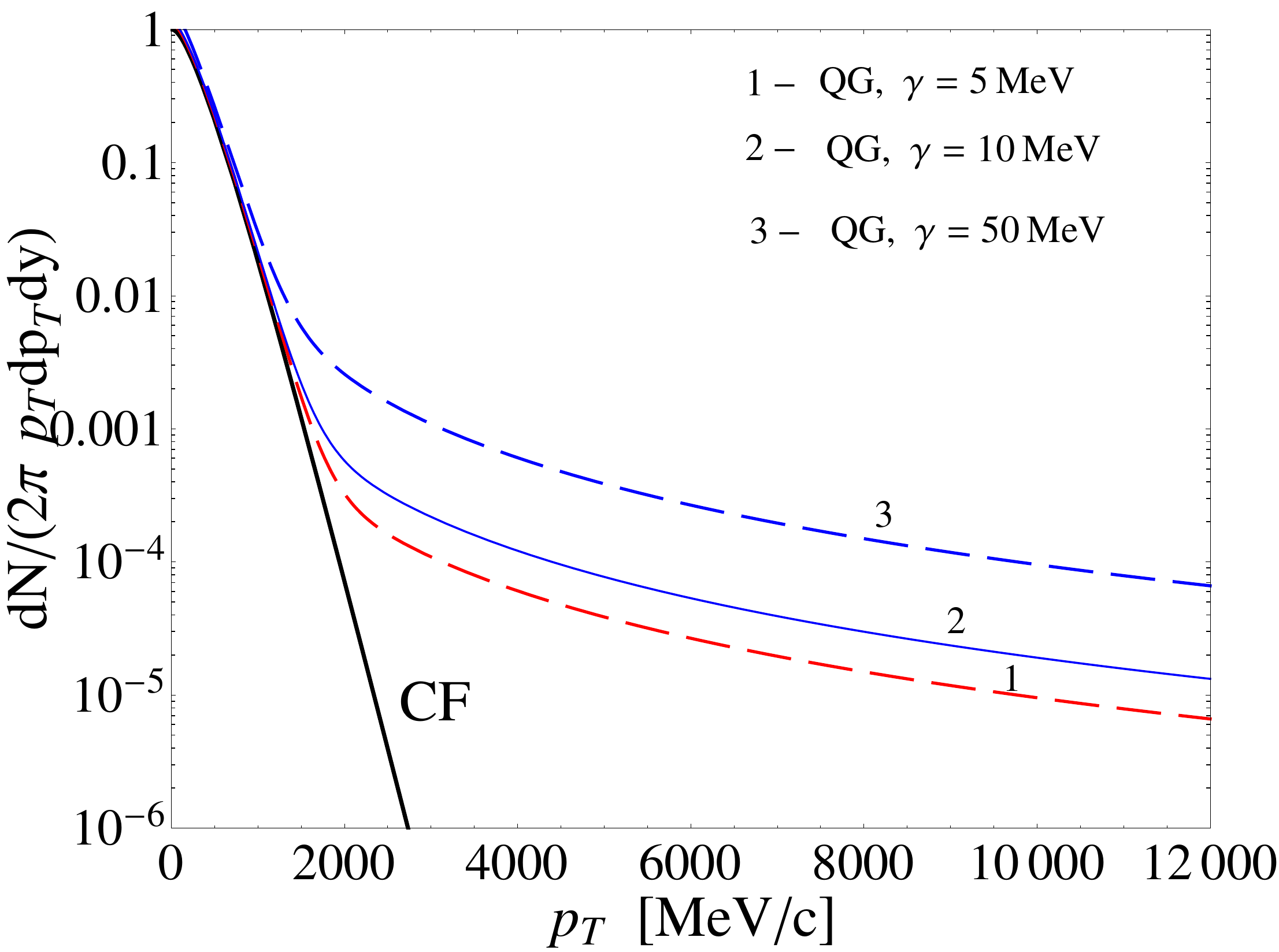}
\caption{ {\it Left panel:} Pion spectrum versus momentum modulus
obtained in two approaches. Black solid curve marked as $f_{\rm BE}$
is the Bose-Einstein distribution at $T = 160$~MeV and $\mu = 0$.
Three upper curves, which represent the quantum generalization
approach, are marked by a value of the parameter $\gamma$. {\it
Right panel:} Pion spectrum versus transverse momentum at
midrapidity.
A straight black line marked as ``CF'' was obtained in the
Cooper-Frye approach. Three upper curves, which represent the
quantum generalization approach, are marked by the digits following
the parameter $\gamma$ value. } \label{fig:bw-spectrum}
\end{figure}

We have calculated the spectrum of pions (\ref{3:109a}), as a
dependence on the $|\bs p| = p$, using a representation of the
element of the momentum space in the form $d^3p = \sin{\theta}
d\theta d\varphi p^2 dp$.
To get the dependence of the spectrum only
on a modulus of the particle momentum $p$, we assume the isotropic
symmetry of the pion radiation and integrate over the angles. Then,
formula (\ref{3:109a}) can be written as
\begin{eqnarray}
\frac{d N}{4\pi p^2dp} \,=\, V \int_0^\infty \frac{dk^0}{\pi}\,
\frac{\left( E_p+k_0 \right)^2}{2E_p}\, \frac{ m\gamma
}{(k_0^2-E_p^2)^2+(m\gamma )^2}\, f_{_{\rm BE}}(k^0)\,.
\label{3:110}
\end{eqnarray}
Our goal is to compare this spectrum with the Cooper-Frye formula
(\ref{3:31}), which for the isotropic symmetry of radiation looks
like
\begin{equation}
\frac{d N}{4\pi p^2 dp} \,=\, V \, f_{_{\rm BE}}(E_p) \,,
\label{app:cf-1}
\end{equation}
where $E_p = \sqrt{m^2 + p^2}$.
In both formulae, we take the unit volume of the pion system, $V = 1$.
To show the quantum generalization we have calculated three patterns
of the life time $\tau = 1/\gamma$ of the pion system, i.e.,
$\gamma = 50,\, 100,\, 150$~MeV.
The spectra are plotted in Fig.~\ref{fig:bw-spectrum}, left panel.
One can see a significant difference in the dependencies
generated by the two approaches.
The Cooper-Frye kinetic approach is
represented as the Bose-Einstein distribution function $f_{_{\rm
BE}}$, which on a logarithmic scale is just a straight line.
Meanwhile, the quantum generalization of the kinetic approach, which
takes into account the finite lifetime of the radiating system
(eq.~(\ref{3:110})), is represented by the curves, resembling power
dependence, rather than an exponential one.

%

Our following example implies a comparison of the pion transverse
spectra at midrapidity in the case of the global thermodynamic
equilibrium.
To calculate the spectrum based on the quantum
generalization of the kinetic approach we use eq.~(\ref{3:110}) and
make the transform of the momentum coordinates as in
Section~\ref{subsec:fin-lt}, $(p_x,p_y,p_z) \to (\varphi, p_T, y)$.
For these conditions the spectrum, which depends now on on the
transverse momentum $p_{_{T}} = |\bs p_{_{T}}|$, looks like
\begin{eqnarray}
\frac{d N}{2\pi p_Tdp_T dy} \,=\, V \int_0^\infty \frac{dk^0}{\pi}\,
\left( m_T + k_0 \right)^2\, \frac{ m\gamma
}{(k_0^2-m_T^2)^2+(m\gamma )^2}\, f_{_{\rm BE}}(k^0)\,,
\label{3:111}
\end{eqnarray}
where $m_T = \sqrt{m^2_\pi + p_T^2}$ and $y = 0$.
For the sake of simplicity, we consider the many-particle system as a gas at
rest (no hydrodynamical flows).
For instance, it is the pion system at
midrapidity in the laboratory frame created in the central Pb+Pb
collision in the collider experiment.
These calculations aim to
compare, on the most transparent basis, the spectrum of pions
emitted by a system with a finite lifetime $\tau = 1/\gamma$ with
the Cooper-Frye formula, which at midrapidity $y = 0$ looks like
\begin{equation}
\frac{d N}{2\pi p_T dp_T} \,=\, V \, m_T(p_T) \, f_{_{\rm
BE}}\big(m_T(p_T)\big) \,. \label{app:cf-10}
\end{equation}
In both approaches, we assume the azimuthal symmetry. The spectra
are plotted in Fig.~\ref{fig:bw-spectrum}, right panel. A straight
black line marked as ``CF'' was obtained in the Cooper-Frye
approach.
Three upper curves, which represent the quantum
generalization approach, are marked by the digits $1,2,3$ following
a value of the parameter $\gamma = 5,\, 10,\, 50$~MeV, respectively.
Both spectra are normalized to unity at $p_T = 0$.
Again, we can see a significant difference in the behavior of curves,
which correspond to different approaches.
The Cooper-Frye kinetic approach leads to a straight line
(on a logarithmic scale).
Meanwhile, the quantum generalization of the kinetic approach, which takes
into account the finite lifetime of the radiating system (eq.~(\ref{3:111})),
is represented by the curves, resembling power dependence at big $p_T$,
rather than an exponential one.

It is interesting to compare our results with pion spectrum versus
transverse momentum at midrapidity in the transverse momentum range
$0.6$~GeV/c $< p_T < 12$~GeV/c measured in Pb-Pb collisions at
$\sqrt{s_{NN}} = 2.76$~TeV \cite{abelev-2014}.
We will see that
dependencies obtained in the quantum generalization approach, which
accounts for the finite lifetime of the system, and which are
depicted in Fig.~\ref{fig:bw-spectrum} on the right panel, have the
same tendencies as the experimental results, or can even fit the
data at big $p_T$.
At the same time, our calculations within the approach developed in
Section~\ref{subsec:spat-inhomog} show that the same behavior has the
spectra of particles radiating from a spatially inhomogeneous system,
for example from a finite-size system.

\section{Conclusion}
\label{sec:conclusion}

In this work, we investigate the propagation of particles (quantum fields)
from the fireball to the detector by solving the corresponding initial-value problem.
With the help of the ``thermal'' Green's function, which describes a system of
many particles created in collisions of relativistic nuclei and determines
the initial values, we are able to express the detected single-particle spectrum.
For the single-particle spectrum we obtain formula (\ref{eq:27-14}), which coincide
with the Cooper-Frye prescription
\cite{cooper-frye-PRD-v10-1974} (as an example we follow the Bose-Einstein statistics)
\[
E_p\, \frac{d N}{d^3p}\, =\, \int d\sigma_\mu(X)\,p^\mu\, f_{_{\rm BE}}(X;p\cdot u)\,.
\]
We obtained formula (\ref{eq:27-14}) in the framework of the following  assumptions:
(1) The many-particle system has a weak inhomogeneity;
(2) The many-particle system is in local thermodynamic equilibrium;
(3) Particles are in free states on the freeze-out hypersurface;
(4) The correlation function $G^<(x_1,x_2)$ differs
significantly from zero only if the differences of the arguments
${\bf x}_1-{\bf x}_2$ are close to zero.

The latter assumption (4) cannot be applied if the system is close to a second-order
phase transition, when the correlation length becomes sufficiently large.
At the same time, as we discussed in the Section~\ref{sec:loc-equil},
this approximation excludes quantum effects.
Based on the same assumptions (1) - (3), taking into account, in addition to
kinetic contributions, also quantum effects, we obtain the formula (\ref{3:28b-6})
for the single-particle spectrum
\[
2E_p\, \frac{d N}{d^3p} \,
=\, \int \frac{d^3k}{(2\pi)^3 2\omega_k}\,
\int_\sigma d\sigma^\mu(x_1)\, d\sigma^\nu(x_2)\, e^{-i(p-k)\cdot (x_1-x_2) }
(p+k)_\mu (p+k)_\nu \, f(X;k)\,,
\]
where $p^0 = E_p = \sqrt{m^2 + \bs p^2}$, $k^0 = \omega_k = \sqrt{m^2 + \bs k^2}$ and
$f(X;k)$ is the distribution function.
In the case of global equilibrium, this equation  reduces to
\[
2 E_p\, \frac{d N}{d^3p}\, =\, \int \frac{d^3k}{(2\pi)^3 2\omega_k}\, f(k) \,
\bigg| \int_\sigma d\sigma^\mu(x) \,  (p+k)_\mu \,e^{-i(p-k)\cdot x }\bigg|^2 \,.
\]
where $p^0 = E_p$ and $k^0 = \omega_k $.

When stationary many-particle system is in spatially inhomogeneous state,
as discussed in the Section~\ref{subsec:spat-inhomog}, we obtain the formula
(\ref{eq:spectrum4}) to calculate the single-particle spectrum in this case.
To reflect finite lifetime of a spatially homogeneous system of many
particles, we propose to use a parametrization of the spectral function
similar to the relativistic Breit-Wigner function.
The emission of particles from a system with finite lifetime was discussed
in Section~\ref{subsec:fin-lt}, where the formula (\ref{3:109a}) was
obtained for calculating the single-particle spectrum in this case.

In order to clarify the differences in the results of the Cooper-Frye
kinetic approach and a proposed quantum generalization,
Section \ref{subsec:applications} presents single-particle spectra obtained
for several specific cases.
The rapidity distribution, momentum, and transverse momentum spectra at
midrapidity were calculated to compare the emitted spectra of pions on
the most transparent basis.

In conclusion, we note that the generalization of the Cooper-Frye formula
(\ref{3:28b-6}), obtained in this article, takes into account quantum
effects arising for various reasons.
First, there is a complete quantum treatment of the emission process that
includes quantum interference, as in the formula (\ref{3:28b-6}).
Second, the radiation from a spatially inhomogeneous system, as in the formula
(\ref{eq:spectrum4}), which, for example, is applied to a system with
a finite spatial volume.
And, of course, taking into account the finite lifetime of a system of
many particles was obtained in the formula (\ref{3:109a}).

In the next article, the author plans to consider the two-particle spectra
arising from relativistic collisions of particles and nuclei.
Meanwhile, the reader can gain basic knowledge in this area
from Refs.~\cite{GKW,boal,heinz99,anch98,chapman94,anch-heinz-2008}.
For example, final-state interaction effects are very important when
considering two-particle spectra (two-particle interferometry).
An introduction to this topic can be found at
Refs.~\cite{anch98,anch-anch-heinz-2006,anch-1996}.

\section*{Acknowledgements}
The author is grateful to I.~Mishustin, L.~Csernai, L.~Satarov, H.~Stoecker for
useful discussions and comments.
The author also thanks the referees for their valuable comments and suggestions
for improving the manuscript.
The work is supported by the National Academy of Sciences of Ukraine
by its priority project "Fundamental properties of matter in the microworld,
astrophysics, and cosmology".

\appendix
\section{Quantum statistics}
\label{sec:appendix}

%
In quantum statistics, the mean value of the operator $\hat{A}$ is defined as:
\[  \left\langle\, \hat{A}\, \right\rangle\,
=\, \frac1Z \, {\rm Tr}\left( \hat{\rho}\, \hat{A} \right)\,
=\, \frac1Z \, \sum_n \left\langle \Psi_n \big|\, \hat{\rho}\, \hat{A}\, \big|
\Psi_n \right\rangle ,
\quad
Z\, =\, \sum_n \left\langle \Psi_n \big|\, \hat{\rho}\, \big| \Psi_n \right\rangle \,,
\]
where $\{\Psi_n(\bs r_1,\bs r_2, \ldots ,\bs r_N)\}$ is the complete set of
many-particle wave functions and $\hat{\rho}$ is the density matrix.

For the local thermal equilibrium the density matrix reduces to the diagonal form
\[ \rho_{nn'}\, =\, \delta_{nn'} e^{E_n/T}\,. \]
Then, the average of the product of two field operators in the grand canonical
ensemble reads
\[ \left\langle\, \hat{\psi}^+(t_2,{\bs r}_2) \, \hat{\psi}(t_1,{\bs r}_1)\,
\right\rangle\,
=\, \frac1Z \, \sum_n \left\langle \Phi_n \Big| e^{-(H - \mu N)/T}\,
\hat{\psi}^+(t_2,{\bs r}_2) \, \hat{\psi}(t_1,{\bs r}_1) \Big| \Phi_n \right\rangle \,.
\]
The lesser Green's function is defined on the base of this averaging
\[
i\hbar\, G^<(x_1,x_2)\,
=\, \pm \, \left\langle\, \hat{\psi}^+(x_2) \, \hat{\psi}(x_1)\, \right\rangle \,,
\]
where $x = (t,\,{\bs r})$, the plus sign reads for bosons and the minus sign
for fermions.

In accordance with the quantum fluctuation-dissipation theorem for the system
which in local (global) thermal equilibrium the lesser Green's function can be
expressed as \cite{kadanoff,chou1985,mrowczynski-1990}
\[
i\hbar\, G^<(\omega,\bs k)\, =\, A(\omega,\bs k)\,f(\omega) \,,
\]
where $G^<(\omega,\bs k)$ is the Fourier transform of the Green's function
$G^<(x_1-x_2)$, $A(\omega,\bs k)$ is the spectral function, $f(\omega)$ is the
Bose-Einstein or Fermi-Dirac distribution function.


\end{document}